\documentclass{iopjournal}

\usepackage{amsmath, amssymb, amsthm, mathrsfs,bm}
\usepackage{hyperref}
\usepackage{placeins}  
\usepackage{graphicx} 
\usepackage[sorting=none]{biblatex}
\usepackage{titling}
\usepackage{fancyhdr}
\usepackage{amsmath} 
\usepackage{booktabs} 
\usepackage{amsfonts} 
\usepackage{mathtools} 
\usepackage{float} 
\usepackage{tabularx}
\usepackage{array}
\usepackage{cancel}
\usepackage[normalem]{ulem}

\newcommand{\redout}[1]{%
  \ifmmode
    \textcolor{red}{\cancel{#1}}%
  \else
    \textcolor{red}{\sout{#1}}%
  \fi
}
\newcommand{\replace}[2]{%
  \redout{#1}\,\textcolor{red}{#2}%
}

\providecommand{\apjl}{The Astrophysical Journal Letters}

\begin{document}

\articletype{Paper} 

\title{A Partially Supervised Latent-Space Variational Autoencoder for Generating Neutron Star Equations of State
}

\author{Tianqi Zhao$^{1,2*}$\orcid{0000-0003-4704-0109}, Fanglida Yan$^3$\orcid{0000-0003-4461-420X}, Alex Ross$^2$\orcid{0009-0005-8103-5823} and James M. Lattimer$^{4}$\orcid{0000-0002-5907-4552}}


\affil{$^1$ Network for Neutrinos, Nuclear Astrophysics, and Symmetries (N3AS), University of California, Berkeley, CA 94720, USA}

\affil{$^2$ Institute for Nuclear Theory, University of Washington, Seattle, WA 98195, USA}

\affil{$^3$ Amazon Science, Seattle, WA 98109, USA}

\affil{$^4$ Department of Physics \& Astronomy, Stony Brook University, Stony Brook, NY 11794, USA}

\affil{$^*$ Author to whom any correspondence should be addressed.}

\email{tianqi.zhao@berkeley.edu}

\keywords{neutron star, equation of state, dense matter, neural network, variational autoencoder}

\begin{abstract}
We develop a {partially supervised latent-space variational autoencoder (PSL-VAE)} framework to reconstruct and generate neutron star (NS) equations of state (EOS). The {PSL-VAE} consists of an encoder network that maps high-dimensional EOS data into a lower-dimensional latent space and a decoder network that reconstructs the full EOS from the latent representation. The latent space includes supervised NS observables derived from the training EOS data, as well as variational latent variables that capture additional EOS features learned automatically.  Using a {PSL-VAE} trained on a Skyrme EOS dataset, we find that a latent space consisting of two supervised observables, the maximum mass $M_{\max}$ and the canonical radius $R_{1.4}$, together with a single variational latent variable associated mainly with the EOS near the crust-core transition, is sufficient to reconstruct Skyrme EOSs with high fidelity. The decoder reconstructed EOSs reproduce $M_{\max}$ and $R_{1.4}$ with mean absolute percentage errors within {$0.08\%$}. Sampling the latent space generates new EOSs that are causal, thermodynamically stable, and consistent with imposed constraints on the supervised observables. The framework therefore provides a compact and physically interpretable parameterization of the NS EOS that is well suited for Bayesian inference with multimessenger observations, including pulsar mass-radius measurements and gravitational wave data.

\end{abstract}

\section{Introduction}


The equation of state (EOS) of cold, beta equilibrated dense matter determines the global structure of a neutron star (NS) \cite{Lattimer2001,Lattimer2007}. Among the many macroscopic quantities that can be derived from a given EOS, two are especially useful because they probe different density regimes and are already constrained by observation: the maximum gravitational mass, $M_{\max}$, and the radius of a canonical $1.4\,M_\odot$ star, $R_{1.4}$. The first is controlled primarily by the pressure at the highest densities reached in stable stars \cite{Lattimer2021,Drischler2021}, whereas the second is governed mainly by the pressure near twice the nuclear saturation density \cite{Lattimer2001}. In this sense, $M_{\max}$ and $R_{1.4}$ provide a compact and physically transparent summary of the high density and intermediate density behavior of the EOS \cite{Margalit2017,Most2018,Bauswein2020,Legred2021,dong2025equation}. This observable-first viewpoint is particularly useful because the underlying microscopic parameterizations are often high dimensional, while the astronomical data constrain only a few combinations of them.

This mismatch is evident in standard EOS parameterizations. In the basic Skyrme model \cite{Bender2003}, the EOS of uniform matter is specified by nine interaction parameters, while relativistic mean-field models are characterized by a similarly high-dimensional set of couplings \cite{Serot1997}. Although these frameworks are useful for forward calculations, their microscopic parameters are not directly observable. Instead, neutron star observations constrain macroscopic stellar properties derived from the EOS, and the underlying interaction parameters must therefore be inferred indirectly.

The goal is to infer the neutron star EOS from observational data. In conventional Bayesian approaches, this requires sampling high dimensional EOS model spaces, such as Skyrme or {relativistic mean-field (RMF)} parameterizations, while repeatedly computing the corresponding EOSs and neutron star observables throughout the inference \cite{char2020bayesian,malik2022relativistic,zhou2023bayesian,scurto2024general,Beznogov2024,huang2025constraining}. {Such model-specific Bayesian analyses infer the EOS conditional on the adopted theoretical model and its parameter prior, and are therefore distinct from approaches designed to explore a broader, more model-agnostic EOS space.} In many other studies, microscopic interaction models with high dimensional parameter spaces are bypassed, and the EOS is instead generated from phenomenological parameterizations. Many such parameterizations are based on Taylor expansions around a reference density, such as the vacuum limit \cite{Tews2017}, the crust-core transition density \cite{Lindblom2010}, or the empirical nuclear saturation baryon density, $0.16$ fm$^{-3}$ \cite{Lattimer2001}. Such constructions can implicitly correlate the low density and high density EOS, since the same expansion coefficients control the behavior over an extended density range. Others use piecewise polytrope \cite{Read2009} or piecewise sound speed \cite{Altiparmak2022}, which limit correlations in the pressure between different densities to within individual segments. Gaussian-process methods model the correlation explicitly with correlation length as a hyperparameter, so that NS EOS can be sampled as a random functional by parameterizing the covariance matrix \cite{Landry2019,Mroczek2023}. In all of these cases, however, the EOS and the corresponding neutron star observables must still be evaluated repeatedly during Bayesian inference. {In the present work, we focus on the former, model-specific setting and seek to replace the high-dimensional parameter space of a given EOS model class with a compact latent representation. Any subsequent Bayesian inference using this representation therefore remains conditional on the EOS model class and training distribution from which it is constructed.} A related line of work seeks to establish a more direct mapping from global stellar observables to the underlying EOS. This has been pursued using empirical correlations based on comprehensive mass-radius curves \cite{Ofengeim2023,cai2023core,sun2025correlations}, which can then be combined with observational data in a later inference step \cite{sun2026new}.

Machine learning methods have recently begun to address this problem. Early supervised approaches trained neural networks to infer dense matter properties or EOS parameters directly from neutron star observables, showing that meaningful information can be extracted from masses, radii, and tidal deformabilities \cite{fujimoto2017methodology,Fujimoto2021,Ferreira2022,Soma2023,Ventagli2025}. More recently, variational autoencoder based approaches have been explored as a way to learn lower-dimensional representations of the EOS itself \cite{han2023nonparametric,Ferreira2025}. These studies demonstrate that neural networks can accelerate EOS inference, but most focus either on the inverse mapping from observables to EOS or on latent representations without explicit supervision by physically transparent neutron star observables.

In this work, we develop a {partially supervised latent-space variational autoencoder (PSL-VAE)} for neutron star EOSs in which the latent space is partially supervised by the maximum mass, $M_{\max}$, and the canonical radius, $R_{1.4}$. The remaining variational latent degrees of freedom encode residual EOS structure not specified by these observables. For the Skyrme EOS ensemble considered here, we find that one variational latent parameter together with $M_{\max}$ and $R_{1.4}$ is already sufficient to reconstruct the EOSs with high fidelity. This yields a three-dimensional latent representation that is much simpler than the original nine-parameter Skyrme space, while remaining directly connected to neutron star phenomenology.

The paper is organized as follows. In Sec.~\ref{subsec:skyrme} and Sec.~\ref{subsec:tov}, we summarize the Skyrme EOS model and the neutron star observables used in this work. In Sec.~\ref{subsec:data}, we describe the construction of the EOS dataset, and in Sec.~\ref{subsec:SSVAE} we introduce the {PSL-VAE} framework and training procedure. The choice of hyperparameters is discussed in Sec.~\ref{subsec:hyperpara}. The main results, including the learned latent space, the dependence of reconstructed EOSs on the latent variables, and the reproduction of known EOSs, are presented in Sec.~\ref{sec:results}. We conclude in Sec.~\ref{sec:conclusion}.

\section{Methods}
\subsection{{Nuclear} Energy Density Functional}\label{subsec:skyrme}
Strongly interacting dense nuclear matter in NSs is often described in the mean-field approximation, in which each nucleon moves in an average potential generated by the surrounding medium. This potential is expressed through an energy density functional constructed from the relevant densities and currents. Two widely used classes of such functionals are the nonrelativistic Skyrme model \cite{Vautherin1972,Chabanat1997,Bender2003} and the relativistic mean-field (RMF) model \cite{Walecka1974,Serot1997}. In this work, we use the Skyrme model to illustrate the {PSL-VAE} framework, since it provides a standard and computationally efficient EOS family with a multi-parameter structure rich enough to test dimensional reduction in latent space. For uniform nuclear matter, spin and current contributions vanish, and the Skyrme energy density takes the form \cite{Tianqi},

\begin{equation}
{\mathcal{E}}_{N} =
\frac{(3\pi^2 n_{n})^{5/3}}{10\pi^{2} m_n^{*}}
+
\frac{(3\pi^2 n_{p})^{5/3}}{10\pi^{2} m_p^{*}}
+
H_{\mathrm{pot}}(n_n, n_p){+n_n m_n+ n_pm_p},
\label{eq:Hsk}
\end{equation}
where the first two terms are the kinetic energy of neutrons and protons in terms of their respective baryon number density, $n_{n/p}$, and effective mass, $m_{n/p}^*$,
\begin{equation}
\begin{aligned}
\frac{m_{n/p}}{m_{n/p}^{*}} &= 1
+ \frac{m_{n/p}}{4}
\Big\{
n_B \left[ t_{1}(2 + x_{1}) + t_{2}(2 + x_{2}) \right]
\\
&\qquad
+ n_{n/p}
\left[ -t_{1}(1 + 2x_{1}) + t_{2}(1 + 2x_{2}) \right]
\Big\},
\end{aligned}
\label{eq:meff}
\end{equation}
and $H_{\mathrm{pot}}$ is the potential energy density given by
\begin{equation}
\begin{aligned}
H_{\mathrm{pot}} &=
\frac{1}{2} n_B^{2} t_{0}
\left( 1 + \frac{x_{0}}{2} \right)
-
\frac{1}{2} (n_n^{2} + n_p^{2}) t_{0}
\left( \frac{1}{2} + x_{0} \right)
\\
&\quad
+
\frac{1}{24} n_B^{\gamma} t_{3}
\left[
n_B^{2} (2 + x_{3})
-
(n_n^{2} + n_p^{2})(1 + 2x_{3})
\right] .
\end{aligned}
\label{eq:Hpot}
\end{equation}
where total baryon number density $n_B=n_p+n_n$.  {In terms of $n_B$ and $\delta\equiv\frac{n_n-n_p}{n_B}$, the saturation density $n_S$ is defined as the density at which the energy per baryon of symmetric nuclear matter is minimized,
\begin{equation}
\left.\frac{P_N(n_B,\delta)}{n_B^2}\right|_{n_B=n_S,\delta=0} = \left.
\frac{\partial}{\partial n_B}
\left(
\frac{\mathcal{E}_N(n_B,\delta)}{n_B}
\right)
\right|_{n_B=n_S,\delta=0}
=0.
\label{eq:saturation_condition}
\end{equation}
Expanding the energy per baryon around symmetric nuclear matter in powers of the isospin asymmetry gives
\begin{equation}
\frac{\mathcal{E}_N(n_B,\delta)}{n_B}
=\frac{\mathcal{E}_N(n_B,0)}{n_B}
+S(n_B)\delta^2+\mathcal{O}(\delta^4).
\end{equation}
Near the saturation density, the symmetry energy can be expanded as
\begin{equation}
S(n_B)=J+L\chi+\mathcal{O}(\chi^2),
\end{equation}
while the energy per baryon of symmetric nuclear matter is expanded as
\begin{equation}
\frac{\mathcal{E}_N(n_B,0)}{n_B}
=\frac{m_n+m_p}{2}+B+\frac{1}{2}K\chi^2+\mathcal{O}(\chi^3),
\end{equation}
where $\chi\equiv(n_B-n_S)/(3n_S)$. The coefficients correspond to the nuclear properties: the binding energy $B$, incompressibility $K$, symmetry energy $J$ and symmetry energy $L$ at saturation density $n_S$.}

The complete Skyrme model involves 9 parameters $x_0$, $x_1$, $x_2$, $x_3$, $t_0$, $t_1$, $t_2$, $t_3$, $\gamma$, excluding the fixed neutron and proton masses, $m_n$ and $m_p${, which is fixed at $m=939$ MeV. Among the 9 Skyrme parameters, the momentum-dependent sector is controlled by $x_1$, $x_2$, $t_1$, and $t_2$ through the effective masses, $m_n^*$ and $m_p^*$. However, as shown in Eq.~(\ref{eq:meff}), these four parameters enter the effective masses only through two independent linear combinations. We therefore replace these four parameters with the neutron effective mass in symmetric nuclear matter,
$m_{S}^{*}\equiv m_n^{*}(n_B=n_S,\delta=0)$,
and the neutron-proton effective mass splitting in pure neutron matter,
$\Delta m_S^{*}\equiv m_n^{*}(n_B=n_S,\delta=1)-m_p^{*}(n_B=n_S,\delta=1)$. Therefore, we can choose arbitrary value for $x_1$ and $x_2$, while the remaining 7 Skyrme parameters, $x_0$, $x_1$, $t_0$, $t_1$, $t_2$, $t_3$, $\gamma$, are determined by requiring the EDF to reproduce the 7 bulk nuclear properties $n_S$, $B$, $K$, $J$, $L$, $m_{S}^{*}$ and $\Delta m_S^{*}$. The sampling ranges adopted for these quantities are listed in Table \ref{tab:skyrme_ranges}. Our final results are insensitive to the detailed choice of the sampling ranges. The complete parameter sets and corresponding bulk nuclear properties are provided in the GitHub repository~\cite{Zhao2026SSVAE}.}
\begin{table}
\caption{{Sampling ranges for the nuclear properties used to determine the Skyrme EDF. The neutron and proton bare masses are fixed to $m_n=m_p=m=939$ MeV.}}
\label{tab:skyrme_ranges}
\centering
\begin{tabular}{ccccccc}
\hline
$n_S$ [fm$^{-3}$]& $B$ [MeV]& $K$ [MeV]& $J$ [MeV]& $L$ [MeV]& $m_{S}^{*}/m$& $\Delta m_S^{*}/m$ \\
\hline
(0.155, 0.165)&(-16.2, -15.8)&(180, 280)&(28, 36)&(0, 150)&(0.7, 1)&(-0.5, 0.5)\\
\hline
\end{tabular}
\end{table}

\section{{Neutron Star Matter}}
Uniform nuclear matter in the core of NS consists of nucleons and electrons. The total energy {density}
 is
\begin{equation}
{\mathcal{E}}(n_B, \delta) = {\mathcal{E}}_N(n_B, \delta) + {\mathcal{E}_e}(n_B, \delta).
\end{equation}
The nucleonic energy density $\mathcal{E}_N$ is described by the Skyrme energy density functional given in Eq.~(\ref{eq:Hsk}). 
The electron contribution $\mathcal{E}_e$ corresponds to a free relativistic electron gas,
\begin{equation}
{\mathcal{E}_{e}(n_{e}) = \frac{m_{e}^4}{8\pi^2}[x_e(2x_e^2+1)(x_e^2+1)^{1/2} - \ln(x_e+(x_e^2+1)^{1/2}) ].}
\end{equation}
{where $m_e$ is electron mass and $x_e=(3\pi^2 n_e)^{1/3}/m_e$}.
Imposing charge neutrality requires $n_e = n_p = (1-\delta)n_B/2$, leading to the expression with respect to $n_B$ and $\delta$.

A cold beta-equilibrated EOS is obtained by minimizing the energy density with respect to the isospin asymmetry at fixed baryon number density,
\begin{equation}
    \frac{\partial \mathcal{E}(n_B,\delta)}{\partial \delta}\bigg|_{n_B,\delta=\delta_{eq}(n_B)}=0. \label{eq:beta_eq}
\end{equation}
This condition determines the equilibrium isospin asymmetry $\delta_{\rm eq}(n_B)$ and therefore the composition of matter. The resulting barotropic EOS is then specified by
\begin{equation}
\mathcal{E}(n_B)=\mathcal{E}\!\left(n_B,\delta_{\rm eq}(n_B)\right),
\qquad
P(n_B)=n_B^2\,\frac{d}{dn_B}E\!\left(n_B,\delta_{\rm eq}(n_B)\right), \label{eq:eos}
\end{equation}
which is equivalently expressed as $P(\varepsilon)$, which determines the neutron star structure through the TOV equations, as discussed in Sec. \ref{subsec:tov}.

%

Although such an EOS formally extends to vanishing density and pressure, it should not be applied beyond the homogeneous core of a neutron star. Uniform nuclear matter becomes unstable to clustering below the crust-core transition density $n_{\mathrm{cc}}$. $n_{\mathrm{cc}}$ can be estimated using the thermodynamic spinodal condition\footnote{A more complete treatment may include density-gradient and Coulomb terms, as in the dynamical method \cite{Ducoin2007}. For Skyrme-type functionals, these finite-size effects generally reduce the crust-core transition density relative to the thermodynamic estimate by about $10\%$ \cite{xu2009locating,ducoin2011core}. Since our goal here is to construct a training set based on the bulk EOS, we use the thermodynamic criterion for simplicity.}, which requires the Hessian matrix of the nuclear energy density with respect to $(n_p,n_n)$ to be positive definite{\cite{Kubis2007}},
\begin{eqnarray}
\det[H] &=&
{\frac{\partial^2 \mathcal{E}_N(n_p,n_n)}{\partial n_p^2}\frac{\partial^2 \mathcal{E}_N(n_p,n_n)}{\partial n_n^2}-\left(
\frac{\partial^{2} \mathcal{E}_N(n_p,n_n)}{\partial n_p \partial n_n}\right)^2} \nonumber\\
&=&{\frac{4}{n_B^2}}\left[\frac{\partial^{2} \mathcal{E}_N(n_B,\delta)}{\partial n_B^{2}}\frac{\partial^{2} \mathcal{E}_N(n_B,\delta)}{\partial \delta^{2}}-\left(
\frac{\partial^{2} \mathcal{E}_N(n_B,\delta)}{\partial n_B\,\partial \delta}{-\frac{1}{n_B}\frac{\partial\mathcal{E}_N(n_B,\delta)}{\partial \delta}}\right)^{2}\right]
> 0 .
\label{eq:stability}
\end{eqnarray}
This condition is evaluated for the barotropic EOS along the beta equilibrium composition, $\delta_{\rm eq}(n_B)$. At density below $n_{\mathrm{cc}}$, the above stability condition {Eq. (\ref{eq:stability})} is violated, which marks the onset of an instability of uniform nuclear matter toward clustering into nuclei. This condition is generally more stringent than the hydrodynamic stability criterion $dP/d\mathcal{E}>0$, which ensures mechanical stability through the positivity of the total derivative as in Eq. (\ref{eq:eos}).

Once the crust-core transition density $n_{\mathrm{cc}}$ is determined, the beta-equilibrated core EOS above $\{n_{\mathrm{cc}},\mathcal{E}_{\mathrm{cc}},P_{\mathrm{cc}}\}$ is combined with the SLy4 crust EOS below the transition to construct a complete EOS extending to the stellar surface. The practical implementation is discussed in Sec.~\ref{subsec:SSVAE}.

\subsection{Neutron Star Observables \label{subsec:tov}}

Neutron stars provide a unique laboratory for probing the EOS of dense matter through their macroscopic observables. Accurate NS masses are obtained from radio timing of binary pulsars, particularly in systems exhibiting measurable Shapiro delay \cite{Demorest2010,Antoniadis2013,Cromartie2020}. Radius information is supplied by X-ray pulse-profile modeling with NICER \cite{Miller2021,Riley2021} and by gravitational-wave observations of binary neutron star mergers, which constrain the tidal deformability and hence the EOS \cite{Abbott2018,Abbott2019}. Relating these observations to the underlying microphysics requires the EOS to be connected to stellar structure through the Tolman-Oppenheimer-Volkoff equations \cite{Tolman,Oppy},
\begin{eqnarray}
    \frac{dP}{dr}&=&- \frac{G(m+4\pi Pr^3/c^2)(\mathcal{E}+P)}{r(rc^2-2Gm)},\label{eq:tov_dpdr}\\
    \frac{dm}{dr}&=& 4\pi r^2 \,\frac{\mathcal{E}}{c^2}\label{eq:tov_dmdr}
\end{eqnarray}
where $m(r)$ is the mass enclosed within radius $r$ and $P(r)$ is the pressure at radius $r$. Our TOV solver uses the \texttt{SciPy} LSODA method \cite{Petzold1983LSODA} to integrate Eq. (\ref{eq:tov_dpdr}) and Eq. (\ref{eq:tov_dmdr}) outward from the stellar center, where
$m=0$, $r=0$, $\mathcal{E}=\mathcal{E}_c$, and $P=P_c$, to the surface
$r=R$, defined by the condition $P(R)=0$, at which point $m(R)=M$. Repeating this process across a range of central pressures $P_c$ produces the full mass-radius (MR) relation, $R(M)$. The EOS $\mathcal{E}(P)$ is required to close the system of equations and uniquely determines the neutron star MR relation. For a given EOS, e.g. SLy4 EOS in Fig. \ref{fig:sly4comp}, there exists a maximum central pressure $P_{\max}$ that yields a stable configuration, corresponding to the maximum neutron star mass $M_{\max}$. The associated central baryon number density $n_{\max}$ and energy density $\mathcal{E}_{\max}$ refer to the same configuration, which can be used as the upper bound of  the EOS as in Tab. \ref{tab:input_data}. Precise measurements of NS masses from radio timing observations have already established a robust lower bound $M_{\max}\gtrsim2\,M_\odot$, placing strong constraints on viable EOS models. Notably, the MR curve is nearly vertical over a broad mass range, indicating that neutron star radii remain approximately constant for canonical masses in the range $1.1\,M_\odot$ to $1.6\,M_\odot$. A complementary and commonly quoted quantity derived from the MR relation is the radius of a $1.4\,M_\odot$ neutron star, denoted $R_{1.4}$. Both $M_{\max}$ and $R_{1.4}$ are used as supervised latent observables during training.

\subsection{Training, Validation{, Selection} and Test Data\label{subsec:data}}
We use the Skyrme model discussed in Sec. \ref{subsec:skyrme} to derive our input EOS training data. {We randomly generate $2\times10^6$ Skyrme parameter sets by uniformly sampling the nuclear properties at saturation within the ranges specified in Table~\ref{tab:skyrme_ranges}. However, many of the sampled Skyrme models are well behaved around saturation density but become unphysical at higher densities. For example, the symmetry energy may become negative, in which case the proton fraction approaches zero and beta equilibrium in Eq.~(\ref{eq:beta_eq}) can no longer be established. We therefore impose the high-density pure-neutron-matter condition,
\begin{equation}
P_N(n_B=15n_S,\delta=1)>600 \textrm{ MeV fm}^{-3}
\end{equation}
This condition also removes a large fraction of EOSs that are too soft to support massive neutron stars. Approximately $1\times10^6$ EOSs satisfy this initial filter. All of these EOSs have crust-core transition densities within $0.02~\mathrm{fm}^{-3}<n_{\mathrm{cc}}<0.12~\mathrm{fm}^{-3}$, although a substantial fraction have negative transition pressures, $P_{\mathrm{cc}}$. We therefore additionally require
\begin{equation}
P_{\mathrm{cc}}>0.1~\mathrm{MeV,fm^{-3}},
\end{equation}
which leaves 685716 EOSs. Their crust-core transition densities are concentrated around $n_{\mathrm{cc}}=0.070\pm0.012~\mathrm{fm}^{-3}$. We then solve for the maximum-mass configuration of each EOS and require causality up to the corresponding central pressure,
\begin{equation}
c_{s,\max}^2=\left.\left(\frac{d\mathcal{E}}{dP}\right)^{-1}\right|_{P=P_{\max}}
\end{equation}
This causality requirement is relevant because the Skyrme EDF is intrinsically nonrelativistic and can become acausal at sufficiently high density. After this filter, 674323 Skyrme EOSs remain and are stored in the GitHub repository~\cite{Zhao2026SSVAE}. Finally, for the PSL-VAE dataset, we retain the 545427 EOSs that support a maximum mass $M_{\max}>1.95,M_\odot$. The $1.95,M_\odot$ threshold provides a small margin below the observationally relevant $2,M_\odot$ scale, allowing EOSs with $M_{\max}>2,M_\odot$ to be generated without requiring extrapolation beyond the training domain.}

The structure of the input data is an array of dimensions $N\times d_x$, where $N{=545427}$\footnote{{We use more than $5\times10^5$ Skyrme EOS samples and find empirically that a training set of approximately $10^5$ samples is sufficient to achieve good model performance.}} is the number of distinct EOS. Among data of dimension $d_x=107$, the first $d_{c_s^2}=101$ columns are comprised of sound speed data, with the sound speed $c_s^2$ given by the following equation:
\begin{eqnarray}
c_{s,i}^2&=&\left(\frac{d\mathcal{E}}{dP}\right)^{-1} _{P=P_i}\label{eq:csdef}\\
P_i &=& \exp\!\left[\ln P_{\mathrm{cc}}  + \frac{i}{d_{c_s^2}-1}\big(\ln P_{\mathrm{max}} - \ln P_{\mathrm{cc}}\big)\right], \label{eq:pressuregrid}
\end{eqnarray}
where $P_i$ defines a logarithmically spaced pressure grid between $P_{\mathrm{cc}}$ and $P_{\mathrm{max}}$. The final $d_b=6$ columns that comprise the input data are boundary conditions that set critical components of the NS structure for each EOS. This includes the baryon number density $n_{\mathrm{cc}}$, energy density $\mathcal{E}_{\mathrm{cc}}$, and pressure $P_{\mathrm{cc}}$ at the crust-core transition calculated from Sec. \ref{subsec:skyrme}, as well as the maximum limits of baryon number density $n_{\mathrm{max}}$, energy density $\mathcal{E}_{\max}$, and pressure $P_{\mathrm{max}}$ which can be reached at the core of the most compact (and massive) stable NS introduced in Sec. \ref{subsec:tov}. The boundary value data used is summarized in Tab. \ref{tab:input_data}.

Since the sound speed is defined in terms of the energy density rather than the mass density, it is expressed in natural units and therefore satisfies $0<c_{s}^2<1$. The $d_b=6$ boundary values are also strictly positive. Therefore, the entire input data array is then log-scaled to make the data distribution more symmetrically peaked. This data is then split into 3 datasets that are used for different purposes during training. The training split is summarized in Tab. \ref{tab:data_splits}. The training set, which contains the majority of the data, is used to update the model parameters by minimizing the cost function during the {PSL-VAE} training procedure. The validation set is used to monitor performance during training, determine the stopping point, and select the best-performing model without biasing the final evaluation. The selection set is not used during optimization or checkpoint selection. Instead, it is used to compare independently trained models with different hyperparameter choices and random initializations, enabling the selection of the final model configuration based on its predictive performance. The test set is completely excluded from model development, including parameter optimization, checkpoint selection, hyperparameter choice and random initializations, and is reserved for the final assessment of model performance and unbiased reconstruction accuracy.

model development, including parameter optimization, checkpoint selection, and hyperparameter tuning, and is reserved solely for the final assessment of model performance and unbiased reconstruction accuracy.

\begin{table}
\caption{VAE boundary-value inputs at the core--crust transition and at the maximum-density endpoint.}
\label{tab:input_data}
\centering
\begin{tabular}{lcc}
\hline
Quantity & Core--crust transition & Maximum \\
\hline
Baryon number density & $n_{\mathrm{cc}}{\in[0.039,0.112] \textrm{ fm}^{-3}}$ & $n_{\max}{\in[0.912,0.141] \textrm{ fm}^{-3}}$ \\
Energy density        & $\mathcal{E}_{\mathrm{cc}}{\in[37.2,106.4] \textrm{MeV fm}^{-3}}$ & $\mathcal{E}_{\max}{\in[1160,1859] \textrm{MeV fm}^{-3}}$ \\
Pressure              & $P_{\mathrm{cc}}{\in[0.1,0.7] \textrm{MeV fm}^{-3}}$ & $P_{\max}{\in[579,1099] \textrm{MeV fm}^{-3}}$ \\
\hline
\end{tabular}
\end{table}

\begin{table}
\caption{Dataset splits used for training and evaluation.}
\label{tab:data_splits}
\centering
\begin{tabular}{llll}
\hline
Split & Size & Purpose & During training \\
\hline
Training   & $\sim$60\% & Update model weights & Yes \\
Validation & $\sim$20\% & Best model checkpoint  selection \& early stopping & Yes \\
{Selection} & {$\sim$10\%} & {Hyperparameter and model selection}  & {No}\\
Test       & $\sim$10\% & Final unbiased evaluation & No \\
\hline
\end{tabular}
\end{table}

\subsection{Reconstructing EOS from Data}

Our primary goal is to generate new candidate NS EOS with the decoder part of the {PSL-VAE} given latent space parameters. The {PSL-VAE} outputs an array of size $d_x=107$,
$\{c_s^2(P_i)\}_{i=0,..,d_{c_s^2}-1}$ on the logarithmic pressure grid defined in Eq. \eqref{eq:pressuregrid}, together with 6 boundary quantities introduced in Sec. \ref{subsec:data} and summarized in Tab. \ref{tab:input_data}. These quantities define the valid thermodynamic range between the crust-core transition and the maximum central density of a stable neutron star.

Starting from Eq. \eqref{eq:csdef}, the energy density is reconstructed as
\begin{equation}
\mathcal{E}(P)
=
\mathcal{E}_{\mathrm{cc}}
+
\int_{P_{\mathrm{cc}}}^{P}
\frac{1}{c_s^2(\ln P')}\, dP',
\label{eq:eps_of_p}
\end{equation}
where $c_s^2(\ln P)$ is obtained by spline interpolation of the discrete PSL-VAE output. The baryon number density is then computed from
\begin{equation}
n_B(P)=n_{\mathrm{cc}}\exp\left[
\int_{P_{\mathrm{cc}}}^{P}
\frac{1}{c_s^2(\ln P')(\mathcal{E}+P')}\, dP'\right],
\label{eq:nb_of_p}
\end{equation}
with boundary condition $n_B(P_{\mathrm{cc}})=n_{\mathrm{cc}}$.

The reconstructed core EOS table $\{n_B(P_i),\mathcal{E}(P_i),P_i\}_{i=0,..,d_{c_s^2}-1}$ is finally matched to the SLy4 crust EOS below the crust-core transition {\cite{douchin2001unified}}. In practice, the two EOS tables are concatenated directly with continuous quadratic interpolation used within the respective crust and core EOS tables. Between the crust-core  transition point, $(n_{\mathrm{cc}},\mathcal{E}_{\mathrm{cc}},P_{\mathrm{cc}})$ and the last point in the crust EOS table with smaller baryon number density, energy density, and pressure, the two EOSs are matched by linearly interpolating $\mathcal{E}(P)$ and $n_B(P)$. This construction ensures continuity of $P$, $\mathcal{E}$, and $n_B$ across the matching region. Since both $P$ and $\mathcal{E}$ increase monotonically between the two matching points, $dP/d\mathcal{E}>0$ is maintained throughout the interpolation interval, ensuring mechanical stability of the matched EOS.

\subsection{The partially supervised latent-space variational autoencoder\label{subsec:SSVAE}}

\begin{figure*}[t]
  \centering
  \includegraphics[width=\linewidth]{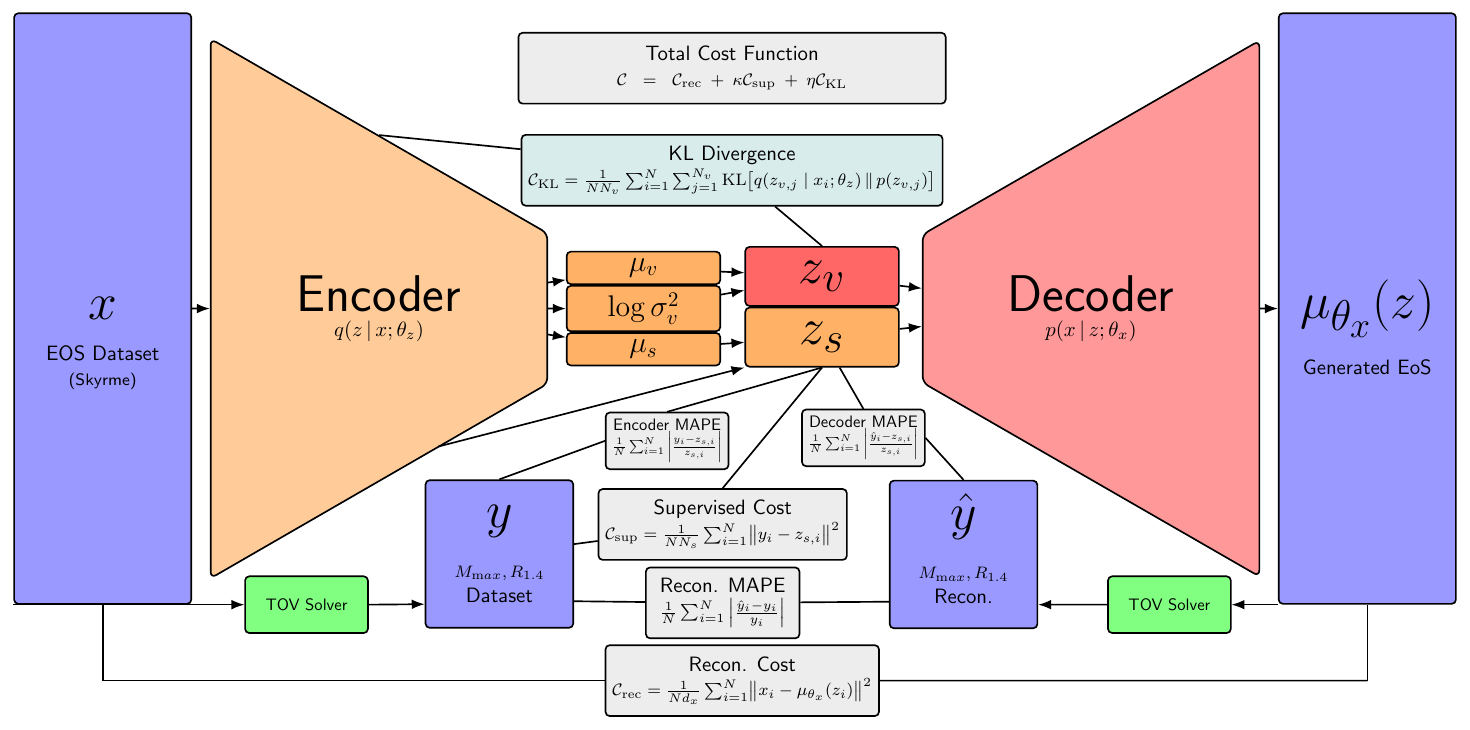}
  \caption{The framework of the partially supervised latent-space variational autoencoder used in this work.\label{fig:VAE}}
\end{figure*}

Figure \ref{fig:VAE} shows the overall framework of the PSL-VAE used in this work, and Tab. \ref{tab:vae_arch} summarizes the network architecture. The encoder maps each EOS input vector $x$ of dimension $d_x=107$ into a latent representation $z$ consisting of a supervised component of dimension $N_s$ and a variational component of dimension $N_v$. The decoder then reconstructs the EOS data vector from latent layer $z$. Consequently, the encoder can be used to predict the observables associated with the supervised component of the latent space, while the decoder can serve as a generator of new EOSs that share the same statistical properties as the EOS data used for training.

Compared with a standard autoencoder, the PSL-VAE adopts a probabilistic latent representation rather than a purely deterministic one. In a conventional autoencoder, the encoder compresses the input into a latent vector that is directly passed to the decoder for reconstruction. In the VAE framework, by contrast, the encoder produces the parameters of a latent distribution, while the decoder models the conditional distribution of the data given the latent variables. This formulation allows stochastic sampling in the latent space through the reparameterization trick, while maintaining differentiability during training. In the present PSL-VAE, the latent space is further separated into supervised and variational sectors, so that selected physical quantities, such as $M_{\max}$ and $R_{1.4}$, are encoded explicitly in the supervised component.

Formally, for a data sample $x$ and latent variable $z$, the generative model is characterized by the prior $p(z)$, the likelihood $p(x\mid z)$, the posterior $p(z\mid x)$, and the marginal likelihood $p(x)$, which satisfy Bayes' theorem,
\[
p(x\mid z)p(z)=p(z\mid x)p(x).
\]
Because the exact posterior is generally intractable, we introduce neural networks to approximate the relevant probability distributions. Given a dataset $X=\{x_i\}_{i=1}^N$ and a chosen prior $p(z)$, our modeling goal is to learn the parameters of the likelihood $p(x\mid z)\approx p(x\mid z;\theta_x)$ and thereby characterize the posterior $p(z\mid x)\approx q(z\mid x;\theta_z)$ defined by neural network parameters $\theta_x$ and $\theta_z$. To approach this, we aim to maximize the marginal likelihood of the data,
\[
\mathcal{J}(\theta_x)=\sum_{i=1}^{N} \log \int p(x_i \mid z;\theta_x)\,p(z)\,dz,
\]
which requires integrating the product of the likelihood and the prior over the latent variables and summing over all data examples. This expression represents the probability (evidence) of the data under the model. Taking the posterior $q(z\mid x;\theta_z)$ as variational distribution, we can rewrite
\begin{eqnarray}
 \mathcal{J}(\theta_x,\theta_z) &=& \sum_{i=1}^{N} \log \int q(z\mid x_i;\theta_z) \frac{p(x_i \mid z;\theta_x)\,p(z)}{q(z\mid x_i;\theta_z)} dz \geq \mathcal{J}_{ELBO} \label{eq:elbo}\\
 \mathcal{J}_{ELBO}&=& \sum_{i=1}^{N} \int q(z\mid x_i;\theta_z) \log\frac{p(x_i \mid z;\theta_x)\,p(z)}{q(z\mid x_i;\theta_z)} dz \nonumber\\
 &=& \sum_{i=1}^{N}
\left[
\int q(z \mid x_i;\theta_z)\,\log p(x_i \mid z; \theta_x)\,dz
-
\mathrm{KL}\!\left(q(z \mid x_i;\theta_z)\,\|\,p(z)\right)
\right] \label{eq:elbo_kl} \\
&=& \sum_{i=1}^{N} j_{i, ELBO} \nonumber
\end{eqnarray}
where Jensen's inequality is used to derive the evidence lower bound (ELBO) in Eq. (\ref{eq:elbo}), and Kullback-Leibler (KL) in Eq. (\ref{eq:elbo_kl}) is defined as $\mathrm{KL}\left(q(z)\|\,p(z)\right)=\int q(z)\log\frac{q(z)}{p(z)}  \,dz$. This objective is jointly optimized with respect to the posterior parameters $\theta_z$ and the likelihood parameters $\theta_x$.






In a vanilla variational autoencoder 
(VAE) \cite{VAE1} \cite{VAE2}, the prior distribution over the latent variables 
is chosen to be a standard normal distribution:
\[
p(z) = \mathcal{N}(0, I).
\]
The approximate posterior distribution is assumed to be Gaussian with a diagonal covariance 
matrix, whose mean and standard deviation are parameterized by a neural network:
\[
q(z \mid x; \theta_z) = \mathcal{N}\bigl(\mu_{\theta_z}(x), \mathrm{diag}(\sigma_{\theta_z}^2(x))\bigr).
\]

In this work, we use a partially supervised latent-space VAE framework (PSL-VAE) where $z=[z_s,z_v]$ consists of a standard variational part, $z_v$ and a supervised part $z_s$. The dimensions of the supervised and variational latent spaces are denoted by $N_v$ and $N_s$, respectively. The prior distribution of the latent variables are,

\begin{eqnarray}
p(z_v) &=& \mathcal{N}(0, I), \\
p(z_s) &=& \mathrm{Uniform}(\mathcal{S}), \quad
\mathcal{S} = \{ z_s \in \mathbb{R}^{N_s} : g(z_s) \le 0 \}.
\end{eqnarray}
Here, $g(z_s)$ encodes non-diagonal boundary constraints defining the feasible region. The uniform prior corresponds to a constant density on $\mathcal{S}$ and zero elsewhere.

The corresponding posterior distributions are,
\begin{eqnarray}
    q(z_v \mid x;\theta_{z_v}) &=& \mathcal{N}\bigl(\mu_{\theta_{z_v}}(x), \mathrm{diag}(\sigma_{\theta_{z_v}}^2(x))\bigr)\\
    q(z_s \mid x;\theta_{z_s}) &=& \mathcal{N}\bigl(\mu_{\theta_{z_s}}(x), \epsilon^2 I\bigr)
\end{eqnarray}
where $\mu_{\theta_{z_v}}(x)$, $\sigma_{\theta_{z_v}}^2(x)$ and $\mu_{\theta_{z_s}}(x)$ are outputs of the encoder network, which consists of a sequence of dense layers, corresponding to Layers 1--4 in Table~\ref{tab:vae_arch}{, and $\epsilon^2$ is a fixed, vanishingly small variance}.

In addition, the PSL-VAE assumes Gaussian likelihood for the data:
\begin{eqnarray}
p(x,y\mid z;\theta_x) &=& p(x \mid z; \theta_x)p(y \mid z_s)\nonumber\\
&=& \mathcal{N}\bigl(x \mid \mu_{\theta_x}(z), \sigma^2_x I\bigr) \mathcal{N}\bigl(y \mid z_s, \sigma^2_y I\bigr), \label{eq:likelihood}
\end{eqnarray}
where the first term on the right-hand side coincides with that of a vanilla VAE, where
$\mu_{\theta_x}(z)$ denotes the output of the decoder network and $\sigma_x^2$ is
typically assumed to be fixed. The second term corresponds to the supervised latent
variable $z_s$, which is constrained by the observed data $y$. Under these assumptions,
the joint log-likelihood can be written as
\begin{eqnarray}
\log p(x,y \mid z; \theta_x)
&=&
-\frac{1}{2\sigma_x^2}\|x - \mu_{\theta_x}(z)\|^2
-\frac{1}{2\sigma_y^2}\|y - z_s\|^2 \nonumber \\
&&
-\frac{d_x}{2}\log(2\pi\sigma_x^2)
-\frac{d_y}{2}\log(2\pi\sigma_y^2),
\end{eqnarray}
where $d_x$ and $d_y=N_s$ denote the dimensionalities of $x$ and $y$, respectively.

Substituting this expression into the ELBO in Eq.~(\ref{eq:elbo_kl}) shows that maximizing
the ELBO is equivalent, up to an additive constant, to minimizing a total cost function
formed by summing the following per-sample contribution over the dataset:
\begin{eqnarray}
-j_{ELBO}
&=&
\frac{
\mathbb{E}_{q(z_v \mid x)}\!\left[\|x - \mu_{\theta_x}(z)\|^2\right]
}{2\sigma_x^2}
+
\frac{\|y - \mu_{\theta_{z_s}}(x)\|^2}{2\sigma_y^2}
\nonumber \\
&&
+
\frac{1}{2}
\sum_{j=1}^{N_v}
\left(
\mu_{\theta_{z_v},j}^2
+
\sigma_{\theta_{z_v},j}^2
-
\log\sigma_{\theta_{z_v},j}^2
-
1
\right).
\end{eqnarray}
where $z=[z_s,z_v]$ with the expected mean squared error achieved by batch sampling of $z_v$ using the reparameterization trick:
\begin{eqnarray}
z_v &=& \mu_{\theta_{z_v}}(x) + \sigma_{\theta_{z_v}}(x) \odot \varepsilon,
\qquad
\varepsilon \sim \mathcal{N}(0, I) \nonumber\\
z_s &=& \mu_{\theta_{z_s}}(x), \label{eq:sampling}
\end{eqnarray}
which enables differentiable gradient-based optimization. The last term corresponds to the KL divergence 
$\mathrm{KL}\!\left(q(z_v \mid x_i;\theta_{z_v})\,\|\,p(z_v)\right)$, which involves two normal distributions and can be written analytically. For the supervised latent variables, the KL divergence between the narrow Gaussian posterior and the continuous uniform prior evaluates to 
\begin{equation}
D_{KL}(q(z_s|x_i;\theta_{z_s}) \,\|\, p(z_s)) = -\frac{N_s}{2} \log(2\pi e \epsilon^2) + \log V_{\mathcal{S}},
\end{equation}
where $V_{\mathcal{S}}$ is the volume of the uniformly supported region $\mathcal{S}$. Because this term depends entirely on the fixed hyperparameter $\epsilon$ and the prior volume, it is completely independent of the neural network's trainable parameters $\theta$. Consequently, its gradient with respect to the network parameters is exactly zero ($\nabla_\theta D_{KL} = 0$), and it rigorously drops out of the optimization objective. This formulation safely avoids infinite singularities and mathematically justifies bypassing the stochastic sampling step for $z_s$ in practice, as sampling from a Gaussian as $\epsilon \to 0$ converges to the deterministic mean $\mu_{\theta_{z_s}}(x_i)$.

Throughout training, a total cost function is constructed to jointly optimize the reconstruction accuracy of the EOS, the supervised agreement between the predicted and true supervised latent observable values, and the Kullback-Leibler divergence that regularizes the latent space toward a standard normal prior, thereby balancing physical fidelity, predictive performance, and latent-space smoothness. The cost function used is given by
\begin{eqnarray}
\mathcal{C}_{\textrm{PSL-VAE}}&=&-\frac{2\sigma_x^2}{N d_x}\mathcal{J}_{ELBO}=\mathcal{C}_{rec}+\kappa\mathcal{C}_{sup}+\eta\mathcal{C}_{KL}
\nonumber\\
&=&
\frac{1}{N\,d_x}
\sum_{i=1}^N
\left\lVert x_i - \mu_{\theta_x}(z_i) \right\rVert^2
\;+\;
\frac{\kappa}{N\,N_s}
\sum_{i=1}^N
\left\lVert y_i - \mu_{\theta_{z_s}}(x_i) \right\rVert^2 \nonumber\\
&&+
\frac{\eta}{N\,N_v}
\sum_{i=1}^{N}
\sum_{j=1}^{N_v}
\left[
-\frac{1}{2}
\left(
1 + \log \sigma_{z_v,i,j}^{2}
- \mu_{z_v,i,j}^{2}
- \sigma_{z_v,i,j}^{2}
\right)
\right],\label{cost}
\end{eqnarray}
where $\mathcal{C}_{rec}$ is the reconstruction cost between the input and reconstructed data, $\mathcal{C}_{sup}$ is the reconstruction cost between the input supervised latent observables and those predicted by the encoder, $\mathcal{C}_{KL}$ is the KL divergence, and $\kappa = \frac{\sigma_x^2 N_s}{\sigma_y^2d_x}, \eta = \frac{2\sigma_x^2 N_v}{d_x}$ are weighting terms. The weighting terms are normalized by $N_v$ and $N_s$, the dimensions of the variational and supervised latent spaces, respectively. In practice, $\eta$ and $\kappa$ are treated as hyperparameters. While their physical meaning can be traced back to the likelihood scales $\sigma_x$ and $\sigma_y$ as defined in the likelihood Eq. (\ref{eq:likelihood}).

\begin{table}
\caption{Neural network architecture.}
\label{tab:vae_arch}
\centering
\begin{tabular}{llcc}
\hline
Layer & Type & Neurons & Activation \\
\hline
Input        & N/A    & $d_x$ & N/A \\
Layer 1      & Dense  & 128 & SiLU \\
Layer 2      & Dense  & 64  & SiLU \\
Layer 3      & Dense  & 64  & SiLU \\
Layer 4      & Dense  & $N_s + 2N_v$  & Linear \\
Latent layer (a): $\mu_{\text{s}}$ & Slice & $N_s$ & Linear \\
Latent layer (b$_1$): $\mu_{\text{v}}$ & Slice & $N_v$ & Linear \\
Latent layer (b$_2$): $\log\sigma^2_{\text{v}}$ & Slice & $N_v$ & Linear \\
Latent layer (b): $z_{\text{v}}$ & Sampling & $N_v$ & N/A \\
Latent layer: $z=[\mu_{\text{s}},z_{\text{v}}]$ & Concatenate & $N_s + N_v$ & N/A \\
Layer 5      & Dense  & 64  & SiLU \\
Layer 6      & Dense  & 64  & SiLU \\
Layer 7      & Dense  & 128 & SiLU \\
Layer 8      & Dense  & $d_x$ & Linear \\
Output (a): $\log c_{\text{s}}^2$   & Slice & $d_{c_{\text{s}}^2}$ & NegSoftplus \\
Output (b): $[n,\mathcal{E},P]_{\text{cc,max}}$   & Slice & $d_b$ & Linear \\
Output       & Concatenate & $d_x$ & N/A \\
\hline
\end{tabular}
\end{table}
\subsection{Network architecture and training}
Above is the theoretical foundation for the partially supervised latent-space variational autoencoder (PSL-VAE), whose architecture is shown in Fig. \ref{fig:VAE}, and summarized in Tab. \ref{tab:vae_arch}. We use the \texttt{SiLU} (Sigmoid Linear Unit) activation function throughout the network. \texttt{SiLU} is a smooth variant of \texttt{ReLU} (Rectified Linear Unit), allowing positive inputs to pass largely unchanged while smoothly suppressing negative inputs toward zero \cite{hendrycks2016gaussian}. The output activation function consists of a \texttt{NegSoftplus} activation function for the first 101 columns, corresponding to $\log c_s^2(p_i)$, to enforce the causality and stability constraints $0< c_s^2(p_i)<1$. We find that varying the output activation function does not significantly affect the results of this work; however, the above choice performs slightly better among the tested options.

The primary machine learning system we use is the Python package \texttt{Tensorflow} \cite{2016arXiv160508695A}. We use a batch size of $N_b=64$ and a learning rate of $\alpha=0.0001$. The input data, consisting of the array described in Sec.~\ref{subsec:data} and the supervised latent observables introduced in Sec.~\ref{subsec:tov}, are standardized by subtracting the mean and scaling by the standard deviation, so that each quantity has zero mean and unit variance. For each training batch, the EOS data are written as $X=\{x_n\}_{n=1}^{N_b}$, where each $x_n$ is a vector of dimension $d_x=107$. Passing $X$ through the encoder yields an output of dimension $(2N_v+N_s)\times N_b$, corresponding to $\mu_{\theta_{z_v}}(X)$, $\log \sigma_{\theta_{z_v}}(X)$ and $\mu_{\theta_{z_s}}(X)$. Applying the stochastic sampling procedure in Eq.~(\ref{eq:sampling}) then gives the latent representation $z=[z_s,z_v]$, which has dimension $(N_v+N_s)\times N_b$. The decoder $\mu_{\theta_x}(z)$ takes a batch of latent variables as input and outputs a matrix of dimension $d_x \times N_b$, which is the same as $X$. The network weights and biases are updated after each training batch. At the end of each epoch, the validation cost is evaluated and compared with its values from previous epochs to monitor the progress of training. If no further improvement in the cost function is observed for five consecutive epochs, the training is stopped. The checkpoint of weights and biases that yield the lowest cost function are restored and adopted for the subsequent analysis.

\begin{figure*}[t]
    \centering
    \includegraphics[width=\textwidth]{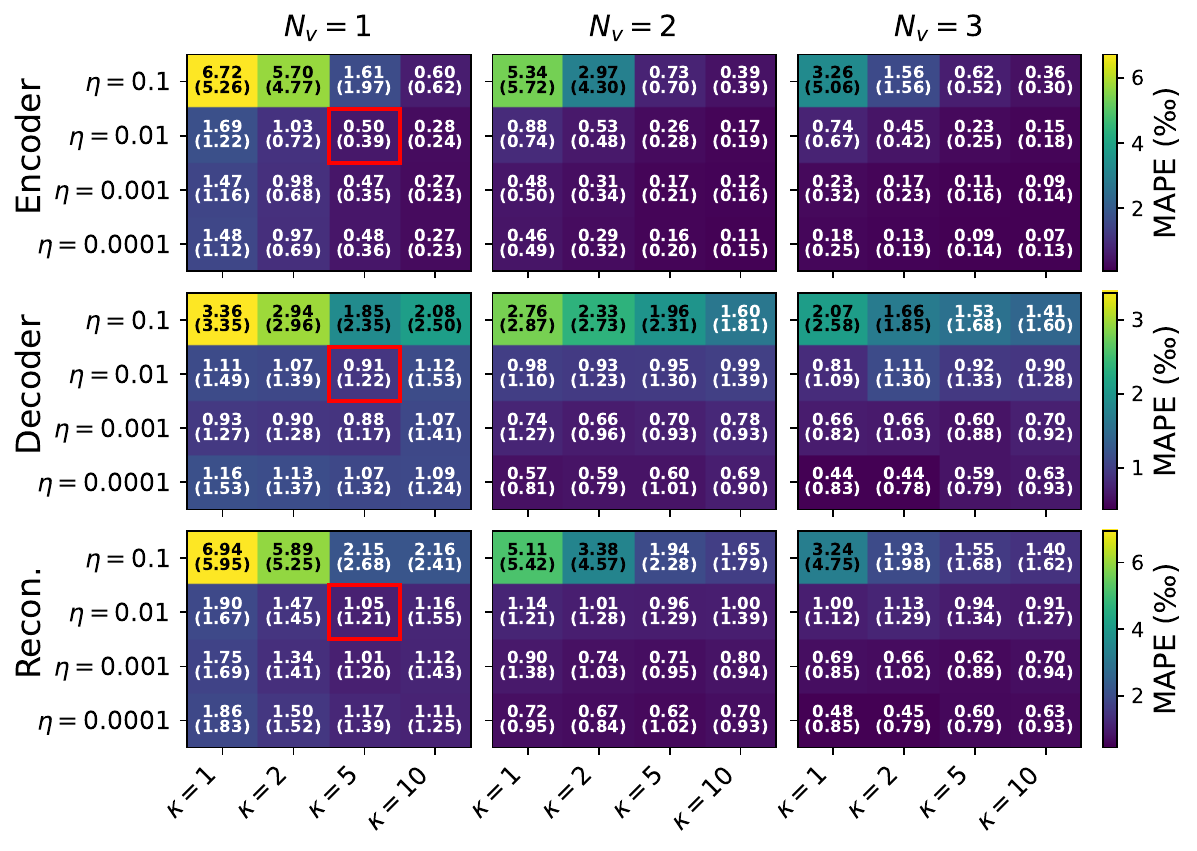}
    \caption{Heatmaps of the MAPE for the first supervised latent observable, $M_{\max}$, averaged over 10 independently trained networks for each combination of the hyperparameters $N_v$, $\kappa$, and $\eta$. The definitions of the encoder, decoder, and reconstruction MAPE are illustrated schematically in Fig.~\ref{fig:VAE}. The corresponding MAPE values for the other supervised latent observable, $R_{1.4}$, show similar qualitative behavior and are indicated in brackets. The hyperparameter choice selected for further analysis is outlined by a red box.}
    \label{fig:Mmax_mape}
\end{figure*}

\subsection{Choice of hyperparameters\label{subsec:hyperpara}}

For each combination of hyperparameters, with $\eta=[0.1,0.01,0.001,0.0001]$, $\kappa=[1,2,5,10]$, and $N_v=[1,2,3]$, we trained 10 PSL-VAE networks independently using different random seeds and initializations. The network performance is evaluated using the supervised observables $M_{\max}$ and $R_{1.4}$. For the selection dataset, we consider three corresponding sets of supervised quantities. First, we solve the TOV equations directly using the selection EOS data to obtain the reference values. These are the same observables as those used for supervision during training, but here they are evaluated from the selection EOSs. Second, we apply the encoder to the selection EOS data, which were not used during training, and obtain the predicted supervised latent quantities. Third, we pass the latent variables through the decoder without stochastic sampling, using only the latent means, to reconstruct the EOS data. Solving the TOV equations for the reconstructed EOSs then gives the corresponding decoder generated values of $M_{\max}$ and $R_{1.4}$. For each supervised quantity, we compute the mean absolute percentage error (MAPE), averaged over all EOSs in the selection dataset, for three comparisons: the encoder MAPE, defined between the reference values and the encoder predictions; the decoder MAPE, defined between the encoder predictions and the values obtained from the reconstructed EOSs; and the reconstruction MAPE, defined between the reference values and the values obtained from the reconstructed EOSs. These quantities are illustrated schematically in Fig.~\ref{fig:VAE}. In this framework, the encoder acts as an approximate TOV solver, providing fast predictions of the supervised observables $M_{\max}$ and $R_{1.4}$ without direct numerical integration, as discussed in Sec.~\ref{subsec:tov}. Accordingly, the encoder MAPE measures the accuracy of the encoder in reproducing these observables. The decoder, on the other hand, serves as an EOS generator conditioned on $M_{\max}$, $R_{1.4}$, and the additional variational latent parameters. The decoder MAPE is therefore particularly important, since it quantifies how well the decoder generated EOSs retain the NS observables.

The heatmaps of these three MAPEs, averaged over the 10 independently trained networks, are shown in Fig.~\ref{fig:Mmax_mape}. The results for $M_{\max}$ and $R_{1.4}$ exhibit qualitatively similar behavior and are therefore presented together to avoid repetition. The heatmaps show that all three MAPEs decrease with increasing $N_v$. This trend is expected, since a larger latent dimension provides more degrees of freedom and therefore allows the model to capture more characteristics of the EOS dataset. However, the improvement obtained by increasing $N_v$ is subdominant, indicating that the EOS dataset can already be described effectively by one variational latent parameter together with two supervised latent parameters.   The variation with $\eta$ is the dominant source of change in the MAPEs, since $\eta$ is proportional to the imposed dispersion $\sigma_x^2$ in the likelihood for the training EOS dataset. A smaller $\eta$ corresponds to treating the EOS dataset as having a smaller intrinsic dispersion, and the limit $\eta=0$ reduces the VAE to a standard autoencoder. Although the accuracy improves as $\eta$ decreases, the posterior distribution of the variational latent parameters becomes less constrained and tends to collapse onto a narrow hypersurface. The impact of $\kappa$ on the three MAPEs is different. The encoder MAPE decreases with increasing $\kappa$, reflecting the larger relative weight of $\mathcal{C}_{\mathrm{sup}}$ in the total cost function. By contrast, the decoder MAPE varies only weakly with $\kappa$. The reconstruction MAPE can be understood as reflecting contributions from both the encoder and decoder errors, and therefore follows roughly the same trend as the encoder MAPE. At the same time, its magnitude is generally close to that of the decoder MAPE, since the decoder MAPE is larger than the encoder MAPE except in the regime of very small $\kappa$ or very large $\eta$.


Based on the MAPE heatmaps, we select a particular set of hyperparameters for further analysis in Sec.~\ref{sec:results}. The adopted values are $\eta=0.01$, $\kappa=5$, and a variational latent dimensionality of $N_v=1$. Although increasing $N_v$ can further reduce the MAPE, the improvement is modest and does not justify the additional model complexity. Likewise, although smaller values of $\eta$ improve the accuracy, they lead to a less regular latent-space structure, as discussed further in Sec.~\ref{subsec:latent}. With $\eta=0.01$ and $N_v=1$ fixed, the choice $\kappa=5$ yields the lowest decoder MAPE. The selected hyperparameter combination is marked by a red box in Fig.~\ref{fig:Mmax_mape}. Among the 10 independently trained PSL-VAE networks for this hyperparameter combination, we select the one with the smallest combined decoder MAPE for $M_{\max}$ and $R_{1.4}$. For the selected network, the decoder MAPE is $0.059\%$ for $M_{\max}$ and $0.074\%$ for $R_{1.4}$, both of which are substantially smaller than the corresponding averages shown in Fig.~\ref{fig:Mmax_mape}.

\section{Results\label{sec:results}}
Having introduced the PSL-VAE framework in Sec.~\ref{subsec:SSVAE} and its hyperparameter selection in Sec.~\ref{subsec:hyperpara}, we now focus on the network with the adopted hyperparameters $\eta=0.01$, $\kappa=5$, and $N_v=1$. As discussed above, this choice provides a good balance between reconstruction accuracy and latent space regularization, while avoiding unnecessary model complexity. Among the 10 independently trained networks for this hyperparameter combination, we select the one with the lowest combined decoder MAPE in $M_{\max}$ and $R_{1.4}$ for the analysis presented in this section based on the selection dataset. Evaluating the selected network on the test dataset yields decoder MAPEs of $0.059\%$ for $M_{\max}$ and $0.075\%$ for $R_{1.4}$, which are nearly identical to those obtained for the selection dataset. Since the test dataset is used neither for training nor for model selection, this agreement indicates that the selected PSL-VAE generalizes well to unseen EOSs and captures the key features of the Skyrme EOS with high accuracy.

In the following, we examine both the structure of the learned latent space and the physical properties of the EOSs generated by the decoder. In Sec. \ref{subsec:latent}, We first analyze the distribution of the supervised observables $M_{\max}$ and $R_{1.4}$, together with the variational latent parameter $z_v$, obtained from the encoder output for the test dataset. In Sec. \ref{subsec:vary_latent}, we then investigate how varying the latent variables affects the reconstructed EOSs generated by the decoder and the corresponding neutron star properties. In this way, we show that the decoder can generate EOSs that preserve the target values of $M_{\max}$ and $R_{1.4}$, while using the remaining latent degree of freedom to encode physically meaningful variations in the EOS.

\begin{figure}[ht!]
    \centering
    \includegraphics[width=0.8\columnwidth]{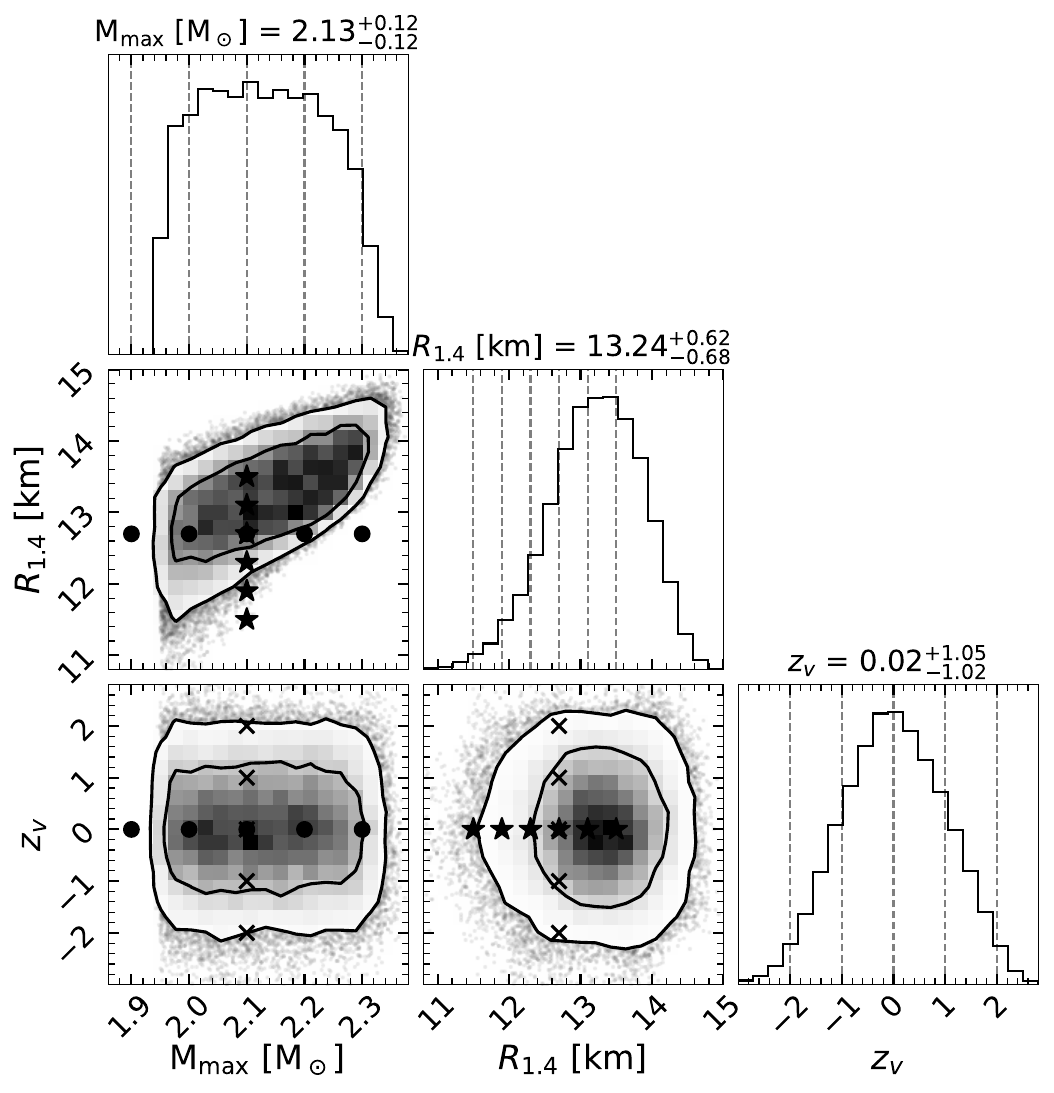}
    \caption{Two-dimensional marginalized distributions of the supervised and variational latent variables for the test dataset. The selected values of $M_{\max}$, $R_{1.4}$, and $z_v$ used to illustrate the latent space sensitivity of the reconstructed EOSs are indicated by dashed lines in the marginalized histograms and by stars, dots, and crosses, respectively, in the two-dimensional panels. The variational latent variable $z_v$ approximately follows a standard normal distribution.}
    \label{fig:latentspace}
\end{figure}

\subsection{Latent space\label{subsec:latent}}


Figure~\ref{fig:latentspace} shows the distribution of the supervised observables $M_{\max}$ and $R_{1.4}$, together with the variational latent parameter $z_v$, obtained from the encoder output for the test dataset described in Sec.~\ref{subsec:data}. Here, $z_v$ denotes the mean of the variational latent variable, $\mu_{z_v}$, and does not include the stochastic sampling introduced by the reparameterization trick in Eq.~(\ref{eq:sampling}). The encoder-predicted values of the supervised observables are very close to the corresponding values obtained by solving the TOV equations directly for the EOS data, consistent with the encoder MAPE being below {$0.05\%$ for $M_{\max}$ and} $0.04\%$ {for $R_{1.4}$}. The occupied region in Fig.~\ref{fig:latentspace} therefore defines the latent parameter space associated with the test EOSs and serves as the basis for generating new EOSs with the decoder.

The condition $M_{\max}>1.95\,M_\odot$ is imposed from observational constraints, leaving a small margin for generating physically relevant EOSs that satisfy $M_{\max}\gtrsim 2\,M_\odot$. The supervised observables exhibit a strong positive correlation, with Pearson correlation coefficient $r(M_{\max},R_{1.4})=0.63$ computed over the full test dataset, as both quantities are sensitive to the stiffness of the EOS. At the same time, they probe different density regimes. The maximum mass is primarily sensitive to the EOS at densities of order four times nuclear saturation density, whereas $R_{1.4}$ is mainly determined by the EOS around twice saturation density \cite{Drischler2021}. As a result, the region with large $R_{1.4}$ and low $M_{\max}$ is excluded by stability requirements, since the pressure must continue to increase with density, corresponding to a positive sound speed squared. Conversely, the region with small $R_{1.4}$ and large $M_{\max}$ is excluded by causality, since the pressure cannot increase too rapidly with density, which requires the sound speed to remain below the speed of light.

The variational latent variable $z_v$ approximately follows a Gaussian distribution centered near zero, consistent with the regularizing effect of the KL divergence. The strength of this regularization increases with the hyperparameter $\eta$. Although smaller values of $\eta$ improve the reconstruction accuracy, the correlations $r(z_v,M_{\max})$ and $r(z_v,R_{1.4})$ increase, making the latent parameter space less regular and therefore less convenient for using the decoder as a controlled EOS generator. We adopt $\eta=0.01$ so that $z_v$ barely correlated with both $M_{\max}$ and $R_{1.4}$, as shown in the lower three panels of Fig.~\ref{fig:latentspace}. This indicates that the variational latent dimension captures a nontrivial degree of freedom in the EOS that is largely independent of the supervised observables.

\subsection{Decoded EOS and Mass-Radius Curves \label{subsec:vary_latent}}
To test the generative capability of the trained PSL-VAE model, we vary individual latent parameters around the central point $M_{\max}=2.1\,M_\odot$, $R_{1.4}=12.7$ km, and $z_v=0$. Each latent parameter is sampled over the supported range shown in Fig.~\ref{fig:latentspace}, while the remaining parameters are held fixed. The resulting sets of latent parameters are decoded into EOSs, which are then used as input to the TOV equations to generate the corresponding MR relations. This procedure allows us to examine whether smooth variations in latent space produce smooth and physically meaningful changes in the EOS and in the corresponding neutron star structure, and therefore whether the decoder provides a stable and interpretable generative mapping from latent parameters to physical observables.

\begin{figure*}[ht]
    \centering
    \includegraphics[width=\textwidth, clip, trim=0 0 0 0]{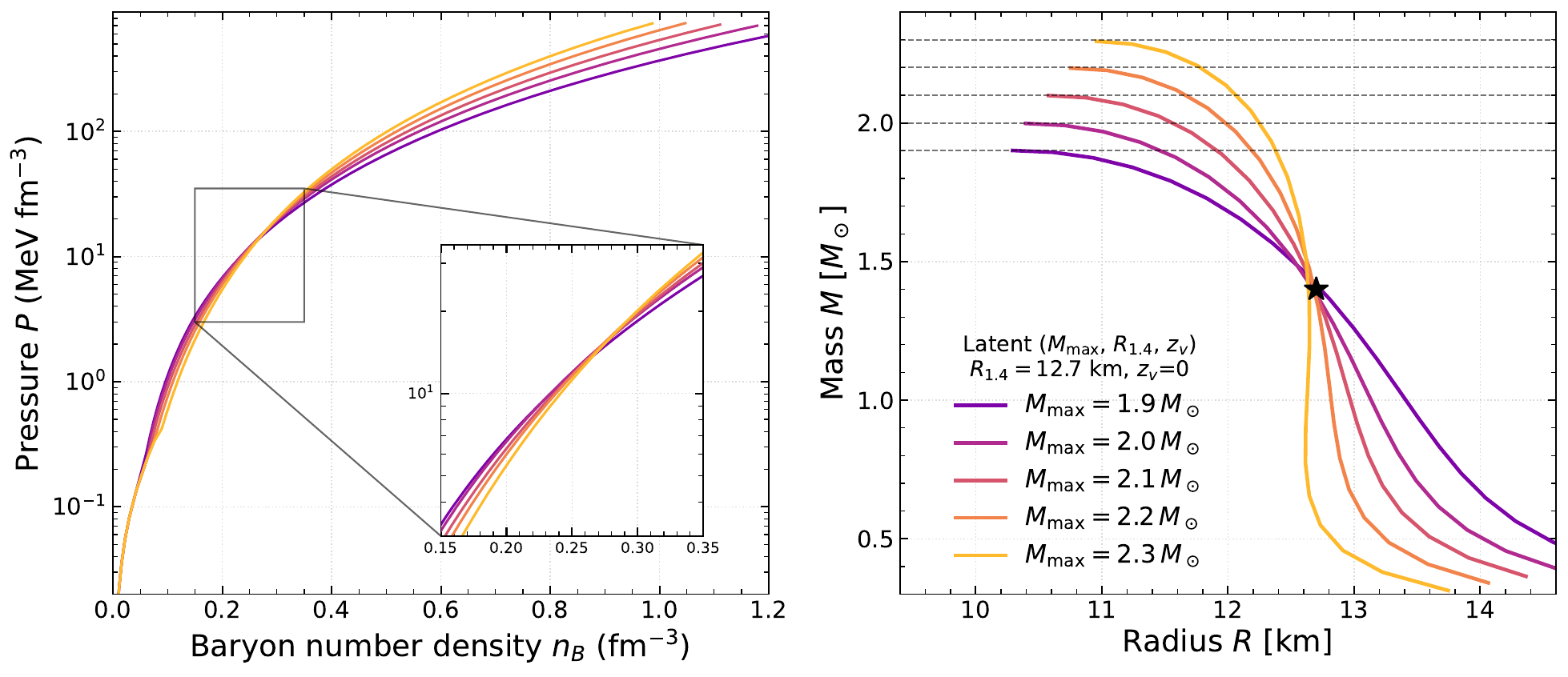}
    \caption{EOSs $P(n_B)$ (left panel) and the corresponding MR relations (right panel) generated by varying the supervised latent observable $M_{\max}$ around the central value indicated by the circular marker in Fig.~\ref{fig:latentspace}. In the right panel, the selected values of $M_{\max}$ are indicated by the horizontal dashed lines. The other latent variables, $R_{1.4}$ and $z_v$, are held fixed, with the fixed value of $R_{1.4}$ marked by the star symbols in the right panel. For each chosen value of $M_{\max}$, the corresponding EOS terminates at a different high density endpoint, determined by the central pressure of the maximum mass configuration.}
\label{fig:EOS_sensitivity_Mmax_P_vs_nB}
\end{figure*}
Figure~\ref{fig:EOS_sensitivity_Mmax_P_vs_nB} shows the sensitivity of the EOS and the corresponding MR relations to variations in $M_{\max}$. We fix the variational latent parameter $z_v=0$ and the supervised latent observable $R_{1.4}=12.7$ km, while varying $M_{\max}$ around the selected central value indicated by circular points in Fig.~\ref{fig:latentspace}. The decoded EOSs remain tightly clustered at intermediate baryon number densities, especially around $n_B\approx 0.25$ fm$^{-3}$, the density range most relevant for determining $R_{1.4}$. This is expected, since fixing $R_{1.4}$ largely constrains the pressure in the vicinity of a few times nuclear saturation density. At higher densities, however, the EOSs separate systematically as $M_{\max}$ is varied. Larger values of $M_{\max}$ correspond to stiffer EOSs at high density, consistent with the need for greater pressure support in the inner core to stabilize more massive neutron stars against gravitational collapse \cite{Lattimer2001}. {This behavior is also consistent with the rough scaling $M_{\max}\propto \replace{\varepsilon}{\mathcal{E}}_{\max}^{-1/2}$ for the maximum-mass configuration, which links a larger supported mass to a stiffer high-density EOS.} Accordingly, the EOSs terminate at different high-density endpoints, which reflect the central baryon density and pressure reached at the maximum-mass configuration for each model. The opposite trend at lower density should not be overinterpreted as an independent physical effect. Because the EOSs in the dataset are smooth functions of density, fixing $R_{1.4}$ while varying $M_{\max}$ forces the trained decoder to redistribute the pressure variation across density in a correlated way. In practice, increasing $M_{\max}$ mainly stiffens the EOS at high density, while inducing only a comparatively small compensating shift at lower density.

\begin{figure*}[ht]
    \centering
    \includegraphics[width=\textwidth]{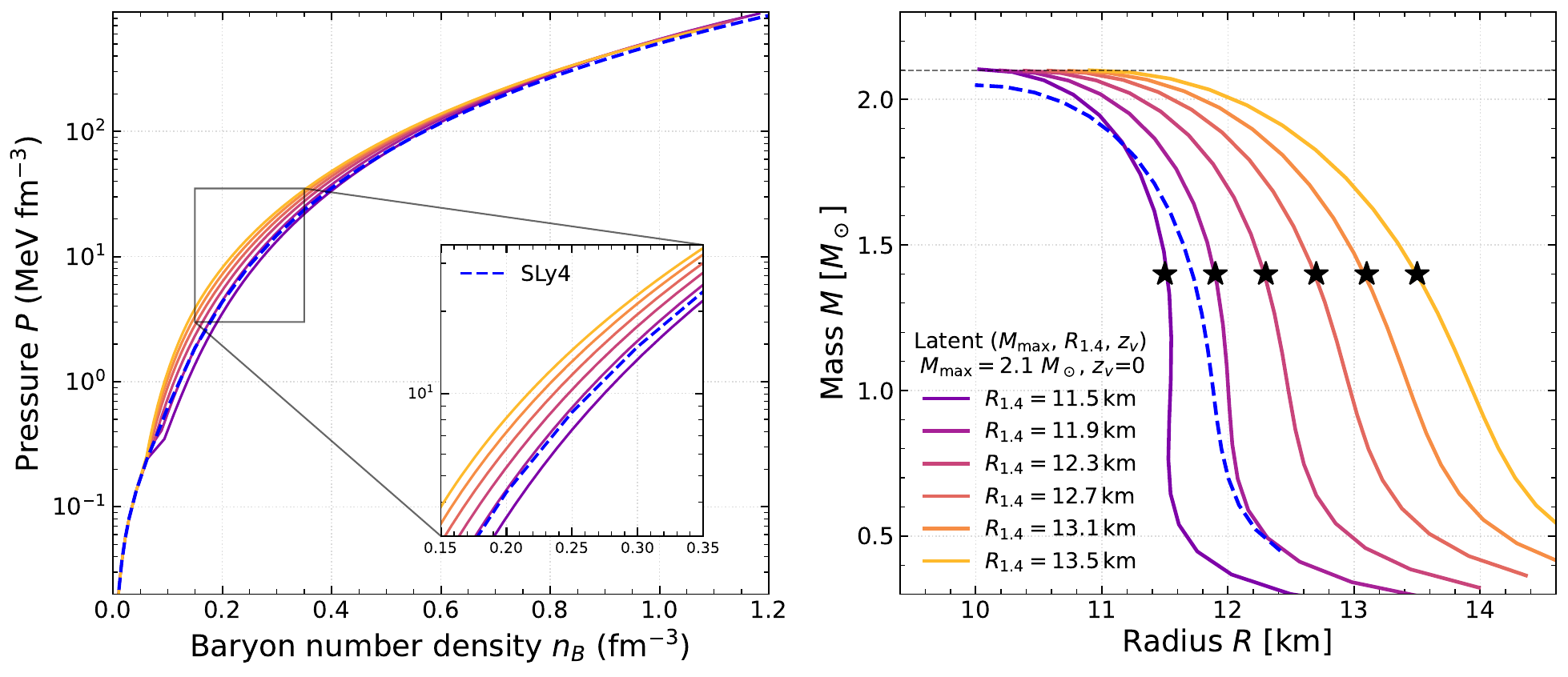}
    \caption{Same as Fig.~\ref{fig:EOS_sensitivity_Mmax_P_vs_nB}, except that the supervised latent observable $M_{\max}$ and the variational latent parameter $z_v$ are held fixed, while $R_{1.4}$ is varied from 11.5 km to 13.5 km. For comparison, the SLy4 EOS is also shown by the dashed blue curve.}
\label{fig:EOS_sensitivity_R14_P_vs_nB_zoom}
\end{figure*}
Figure~\ref{fig:EOS_sensitivity_R14_P_vs_nB_zoom} shows the same analysis as in Fig.~\ref{fig:EOS_sensitivity_Mmax_P_vs_nB}, but now with $M_{\max}$ and $z_v$ held fixed while the supervised latent observable $R_{1.4}$ is varied in steps of 0.4 km. In this case, the decoded EOSs remain relatively similar at high densities, corresponding to the inner core region, but separate clearly in the lower-density range highlighted by the zoomed panel on the left. This behavior is consistent with the well-known sensitivity of neutron star radii to the pressure near and somewhat above nuclear saturation density \cite{Lattimer2004}, where variations in the EOS primarily affect the outer core and envelope rather than the central high-density region. For reference, the exact SLy4 EOS and its corresponding MR relation are also shown. The SLy4 EOS has $M_{\max}=2.05\:M_\odot$ and $R_{1.4}=11.7$ km, and therefore lies between the generated EOSs with $R_{1.4}=11.5$ km and $R_{1.4}=11.9$ km. Compared with these generated EOSs, SLy4 exhibits slightly higher pressure at low density and slightly lower pressure at high density, which is consistent with its smaller maximum mass relative to the fixed value $M_{\max}=2.1\:M_\odot$ adopted for the generated EOSs.

\begin{figure*}[ht!]
    \centering
    \includegraphics[width=\textwidth]{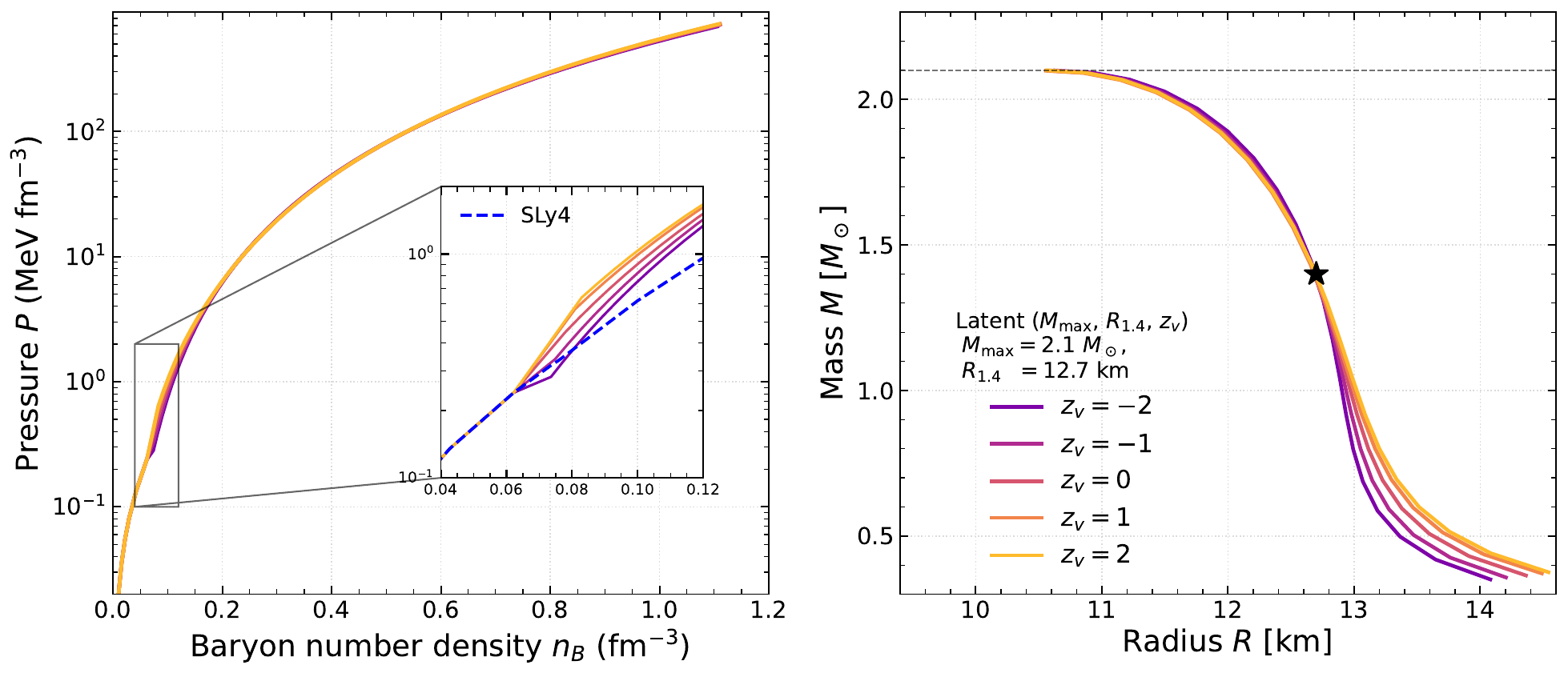}
    \includegraphics[width=\textwidth]{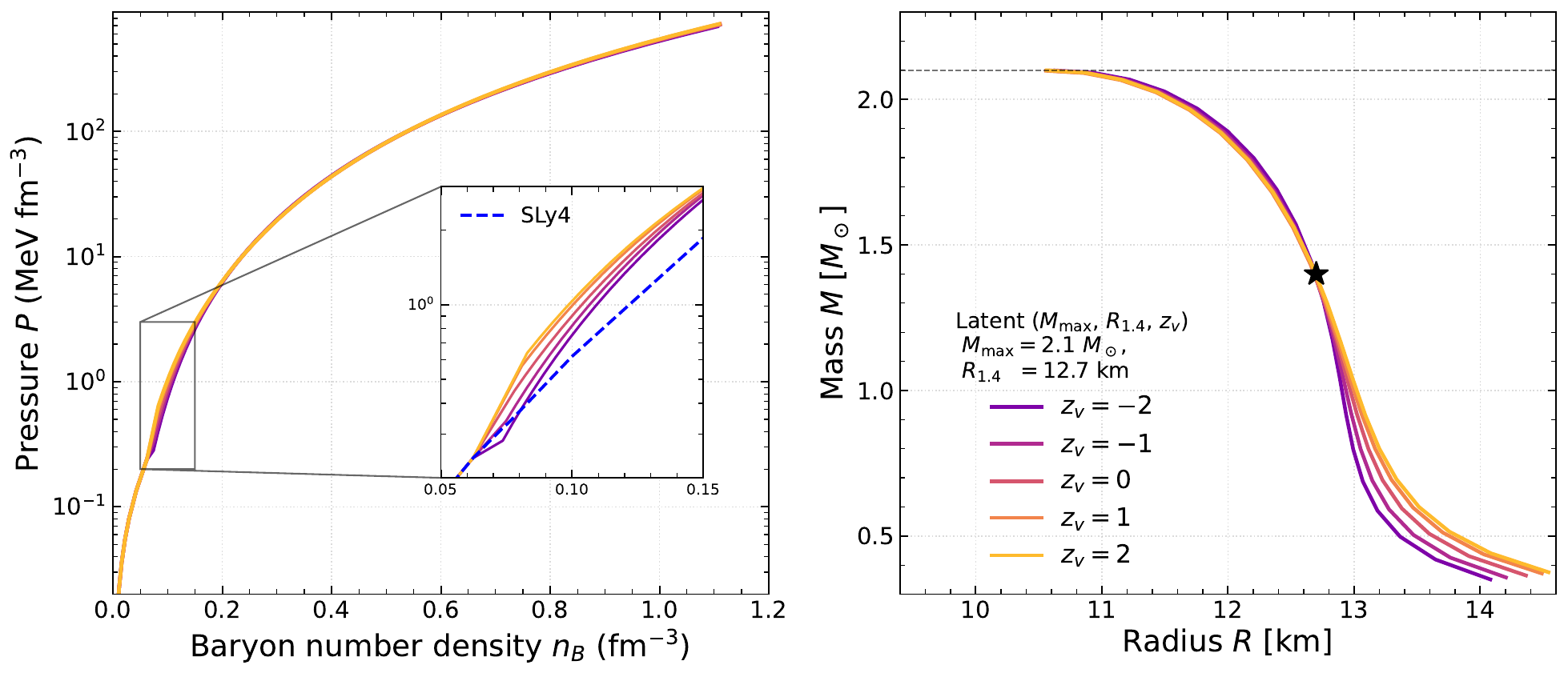}
    \caption{Same as Figs.~\ref{fig:EOS_sensitivity_Mmax_P_vs_nB} and \ref{fig:EOS_sensitivity_R14_P_vs_nB_zoom}, except that the supervised latent observables $M_{\max}$ and $R_{1.4}$ are held fixed, while the variational latent parameter $z_v$ is varied from $-2$ to $2$.}
\label{fig:EOS_sensitivity_variational}
\end{figure*}
We perform the same analysis in Fig.~\ref{fig:EOS_sensitivity_variational}, but now vary the variational latent parameter $z_v$ while holding $M_{\max}$ and $R_{1.4}$ fixed. The resulting EOSs are very similar overall, with the main differences appearing at baryon densities $n_B \lesssim 0.16~\mathrm{fm}^{-3}$. These low density variations translate primarily into differences in the radii of low mass neutron stars, especially for masses $M \lesssim 1\,M_\odot$. For the selected network, $z_v$ is positively correlated with the crust-core transition density and with the pressure in the density range between the crust-core transition and $n_B \lesssim 0.16~\mathrm{fm}^{-3}$. Consequently, larger values of $z_v$ correspond to larger radii for low mass neutron stars. Across different training realizations, the same latent degree of freedom continues to act mainly on the low-density EOS, but the sign of the mapping is not fixed. In some networks, larger $z_v$ produces larger low-mass radii, whereas in others it produces smaller low-mass radii, even though the reconstruction accuracy is comparable. This indicates that the mapping between $z_v$ and specific physical properties is not unique. Different trained networks may represent the same latent degree of freedom through a nonlinear reparameterization, or simply through a reversal of sign. This ambiguity reflects the lack of uniqueness of the latent representation learned by a nonlinear neural network. {We emphasize that the physical manifestation of $z_v$ near the crust-core transition is a straightforward and expected consequence of our specific model constraints. Because the Skyrme parameterization has limited degrees of freedom, and the high- and intermediate-density regimes are already heavily constrained by fixing $M_{\max}$ and $R_{1.4}$ respectively, the remaining structural freedom naturally concentrates in the unconstrained transition region. Observing this expected behavior as $z_v$ varies serves as an important sanity check, validating that the machine learning model has been trained properly and behaves consistently with known physics. In more general, nonparametric EOS families with larger degrees of freedom, it may not always be possible to map learned unsupervised latent variables to such cleanly interpretable physical quantities.}

All three figures, Figs.~\ref{fig:EOS_sensitivity_Mmax_P_vs_nB}--\ref{fig:EOS_sensitivity_variational}, demonstrate that the supervised latent observables $M_{\max}$ and $R_{1.4}$ can be reproduced faithfully by the EOSs generated by the decoder. Some of the latent parameter sets considered here even extend beyond the range covered by the training data, for example when varying $M_{\max}$ to $1.9\,M_\odot$ and $2.3\,M_\odot$ while fixing $R_{1.4}=12.7$ km, or varying $R_{1.4}$ to 11.5 km while fixing $M_{\max}=2.1\,M_\odot$. This indicates that the model has some capacity to extrapolate and generate new EOSs beyond the parameter range represented in the training set. However, the accuracy with which the generated EOSs reproduce the target supervised observables lowers in this extrapolating regime.

\subsection{Decoded reproduction of known EOS}
We now show that the PSL-VAE framework can accurately reproduce known EOSs by directly specifying the supervised latent observables corresponding to those models. As representative examples, we consider the SLy4 EOS {\cite{douchin2001unified}}, for which $M_{\max}=2.05\,M_\odot$ and $R_{1.4}=11.7$ km, and the SkI4 EOS {\cite{reinhard1995nuclear}}, for which $M_{\max}=2.19\,M_\odot$ and $R_{1.4}=12.6$ km. Since the decoder MAPE for the supervised observables is within $0.08\%$, as discussed in Sec.~\ref{subsec:hyperpara}, we treat the target values of $M_{\max}$ and $R_{1.4}$ as the means of the corresponding latent distributions and adopt $0.08\%$ of each target value as the associated standard deviation. For the variational latent parameter $z_v$, we pass the original EOS data through the encoder and obtain the corresponding $\mu_{\theta_{z_v}}$ and $\sigma_{\theta_{z_v}}$ for SLy4 and SkI4. The specific numerical value of $z_v$ is not itself physically meaningful, since the learned variational latent parameter is not unique and varies across different training realizations. Using the means and variances of $M_{\max}$, $R_{1.4}$, and $z_v$, we then generate ensembles of decoded EOSs designed to reproduce the original SLy4 and SkI4 EOSs.

\begin{figure}[ht!]
    \centering
    \includegraphics[width=\textwidth, clip, trim=0 0 0 0]{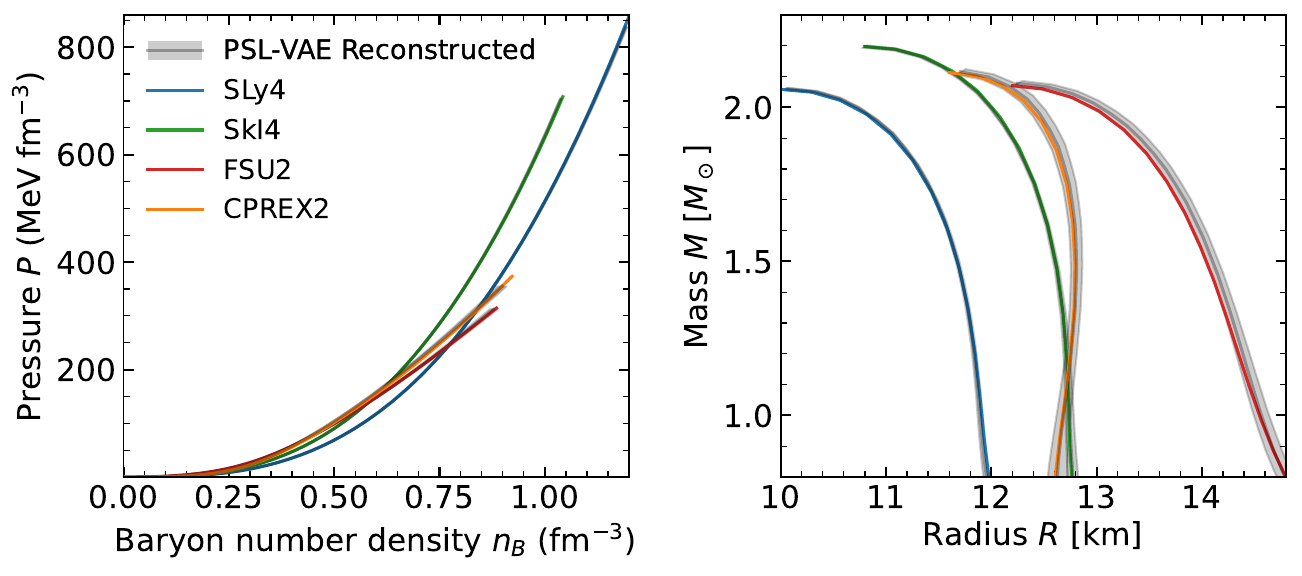}
    \caption{EOSs $P(n_B)$ and the corresponding MR relations generated by fixing $M_{\max}$ and $R_{1.4}$ to the target values of SLy4 and SkI4 { for the Skyrme network, and FSU2 and CPREX2 for the RMF network}, and $z_v$ to the corresponding encoder inferred latent distributions, then decoding the EOSs. The shaded bands denote the $1\sigma$ spread of ensembles generated the decoder. The original SLy4, SkI4{, FSU2 and CPREX2} EOSs and their corresponding MR curves are shown for comparison.}
    \label{fig:sly4comp}
\end{figure}

Figure~\ref{fig:sly4comp} compares the original EOSs and their corresponding MR relations with the ensembles of EOSs generated by the decoder for SLy4 and SkI4. The bands of decoder-generated EOSs represent the $1\sigma$ contours, while the means are indicated by the grey curves. The discrepancy between the decoded EOSs and the original EOSs is barely visible. Minor differences appear in the low-mass part of the SkI4 MR relation, arising from an approximately $4\%$ deviation in the crust-core transition pressure of the decoded EOSs. However, the radii of low-mass neutron stars are extremely sensitive to the {pressure at} crust-core transition, so even a modest shift can produce a visible deviation in that regime. The accuracy of the crust-core transition pressure could likely be improved by introducing an additional hyperparameter that separately weights the reconstruction losses of the sound speed array and the boundary quantities. In terms of the likelihood model, this would correspond to introducing separate dispersions, $\sigma_{x,c_s^2}$ and $\sigma_{x,d}$, for the first $d_{c_s^2}=101$ components and the last $d_b=6$ components, respectively, instead of using a single universal $\sigma_x$ in Eq.~(\ref{eq:likelihood}).

{To further validate the applicability of the PSL-VAE framework to other classes of dense-matter EOSs, we repeat the training procedure using a dataset generated from RMF models, with details provided in Appendix~\ref{app:RMF}. Figure~\ref{fig:sly4comp} demonstrates the ability of the selected RMF-trained PSL-VAE network to reproduce representative RMF EOSs. As one example, FSU2~\cite{chen2014building} is an RMF parametrization calibrated to properties of finite nuclei and neutron stars, including the $2,M_\odot$ maximum-mass constraint. It predicts a relatively large neutron-star radius, $R_{1.4}=14.2$ km. As a second example, CPREX2~\cite{Tianqi} is a recently proposed parametrization constructed to reproduce the parity-violating electron-scattering measurements of $^{48}$Ca and $^{208}$Pb while remaining consistent with known neutron-star mass and radius constraints. Both EOSs are accurately reproduced by the RMF-trained PSL-VAE. The broader shaded bands relative to those obtained for the Skyrme models reflect the larger decoder MAPE of approximately $0.2\%$ for the RMF-trained network, which is adopted as the standard deviation of the supervised latent variables when generating the reconstructed EOS ensembles.}


\section{Conclusion and outlook \label{sec:conclusion}}
In this work, we developed a partially supervised latent-space variational autoencoder framework for neutron star EOSs trained on Skyrme models. The original nine-parameter Skyrme description is reduced to three latent quantities: two supervised observables, $M_{\max}$ and $R_{1.4}$, and one variational latent parameter, $z_v$. For the selected model, with $\eta=0.01$, $\kappa=5$, and $N_v=1$, the decoder MAPE is $0.059\%$ for $M_{\max}$ and $0.075\%$ for $R_{1.4}$. This low-dimensional representation is sufficient to generate EOSs that remain smooth, causal, and thermodynamically stable over the density range relevant for neutron stars.

The three latent quantities were found to control distinct parts of the EOS. Varying $M_{\max}$ mainly changes the high-density stiffness $n_B\gtrsim0.6~\mathrm{fm}^{-3}$and the upper part of the MR relation. Varying $R_{1.4}$ controlling the radii of canonical-mass stars mainly affects the pressure around $n_B\approx 0.25~\mathrm{fm}^{-3}$. At fixed $M_{\max}$ and $R_{1.4}$, varying $z_v$ produces comparatively small changes, concentrated mostly below $n_B\lesssim 0.16~\mathrm{fm}^{-3}$ and most visible in the radii of low-mass stars. The learned latent space is also well behaved: the supervised observables follow the same joint distribution as the training data, while the selected choice $\eta=0.01$ keeps $z_v$ only weakly correlated with them and approximately Gaussian.

The decoder also reproduces known EOSs with high fidelity. For example, SLy4 and SkI4 are both recovered with discrepancies that are barely visible in the EOS and MR relations. The main residual difference appears in the low-mass part of the SkI4 MR curve and is associated with an approximately $4\%$ deviation in the crust-core transition pressure. The latent-space tests further show that the model retains some extrapolating capability. For example, it can generate EOSs with $M_{\max}=1.9\,M_\odot$ and $2.3\,M_\odot$ at fixed $R_{1.4}=12.7$ km, although the training data span only the range $1.95\,M_\odot < M_{\max} < 2.27\,M_\odot$. In this regime, the fidelity with which the target observables are reproduced gradually decreases.

These results show that the PSL-VAE provides a compact surrogate for the original Skyrme parameter space. Instead of exploring 9 microscopic parameters directly, one may work with 3 latent quantities that are much closer to neutron star observables. This makes the decoder a promising tool for model-specific Bayesian inference. Once trained, it can generate large ensembles of EOSs at negligible cost and may therefore be combined with Markov chain Monte Carlo sampling to infer posterior distributions for $M_{\max}$, $R_{1.4}$, and the variational latent degree of freedom from multimessenger data. Such an inference remains conditional on the EOS model class and parameter distribution represented by the training dataset, rather than providing a model-independent inference over all possible NS EOSs.

There are several natural directions for future work. A broader training set could include relativistic mean-field EOSs, chiral effective field theory based constructions, and hybrid hadron-quark EOSs {including crossover transitions such as quarkyonic matter}. We further apply the PSL-VAE framework to 9-parameter relativistic mean-field models and find similar behavior, although two variational latent parameters are needed to achieve comparable accuracy, as discussed in Appendix~\ref{app:RMF}. A third supervised observable such as $R_{2.0}$ {or $R_{\max}$} \cite{Drischler2021,Legred2021,lin2024indication} might be included for more flexible EOS models. Besides, the reconstruction of boundary quantities could also be improved by replacing the single likelihood scale $\sigma_x$ with separate dispersions for the sound-speed array and the boundary data, namely $\sigma_{x,c_s^2}$ for the first $d_{c_s^2}=101$ components and $\sigma_{x,d}$ for the last $d_b=6$ components. Finally, the present study focuses on smooth EOSs. Extending the framework to first-order phase transitions, where Maxwell constructions can produce vanishing sound speed and discontinuous structure \cite{Lopes_2021,Blacker_2023}, will provide a more stringent test of how broadly this latent representation can describe dense matter.

\section*{Code and data availability}

The implementation of the partially supervised latent-space variational autoencoder (PSL-VAE), together with training scripts, preprocessing routines, and example datasets used in this work, is publicly available at GitHub\cite{Zhao2026SSVAE}.

\funding{
T. Z. and A. R. acknowledge the helpful discussion with Sanjay Reddy during Institute for Nuclear Theory Undergraduate Research Network (INTURN) program. T. Z. and A. R. are supported by N3AS's National Science Foundation award No. 2020275. A. R. is supported by  a scholarship from the Mary Gates Endowment for Students. J. M. L. acknowledges support from Grant DE-FG02-87ER40317 from the U.S. Dept. of Energy.}

\appendix
\section{{Additional results with RMF EOS dataset}\label{app:RMF}}

{The PSL-VAE framework is not restricted to NS EOSs modeled with Skyrme EDFs. We repeat the analysis using an EOS dataset based on RMF EDFs. The theoretical framework and implementation of the RMF EDF follow our previous work~\cite{Tianqi}. The EOS dataset, with $d_x=107$, together with the model parameters for $1.8\times10^5$ samples, is available in the GitHub repository~\cite{Zhao2026SSVAE}.}

\begin{figure}[]
    \centering
    \includegraphics[width=0.8\columnwidth]{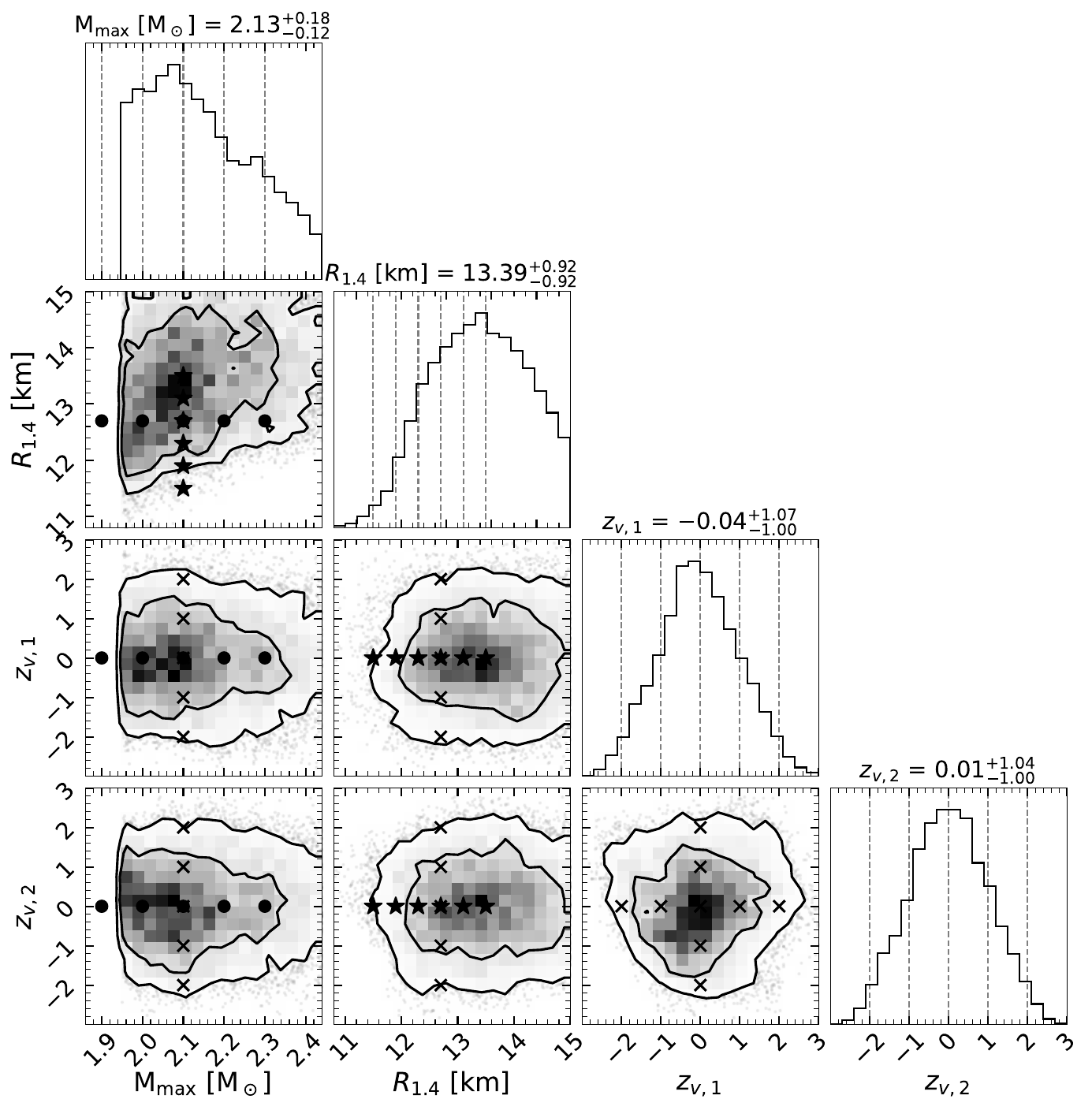}
    \caption{{Same as Fig. \ref{fig:latentspace} but for RMF dataset. The selected PSL-VAE network for RMF dataset has two variational latent parameters, $z_{v,0}$ and $z_{v,1}$.\label{fig:latentspace_rmf}}} 
\end{figure}
{The choice of PSL-VAE hyperparameters is the same as that for the Skyrme dataset, except that the number of variational latent parameters is increased to $N_v=2$ to achieve satisfactory performance. This is not surprising, since the 9-parameter RMF model has more effective degrees of freedom than the 9-parameter Skyrme model, for which only 7 parameters are independent, as mentioned in Sec.~\ref{subsec:skyrme}. The four-dimensional latent space, consisting of two supervised variables ($M_{\max}$ and $R_{1.4}$) and two variational latent parameters ($z_{v,1}$ and $z_{v,2}$), shown in Fig.~\ref{fig:latentspace_rmf}, can capture the majority of the features of the RMF EOSs. The corresponding decoder MAPEs for the test dataset, which is never used in either training or model selection, are $0.196\%$ for $M_{\max}$ and $0.177\%$ for $R_{1.4}$.}

\begin{figure}[ht!]
    \centering
    \includegraphics[width=0.8\columnwidth]{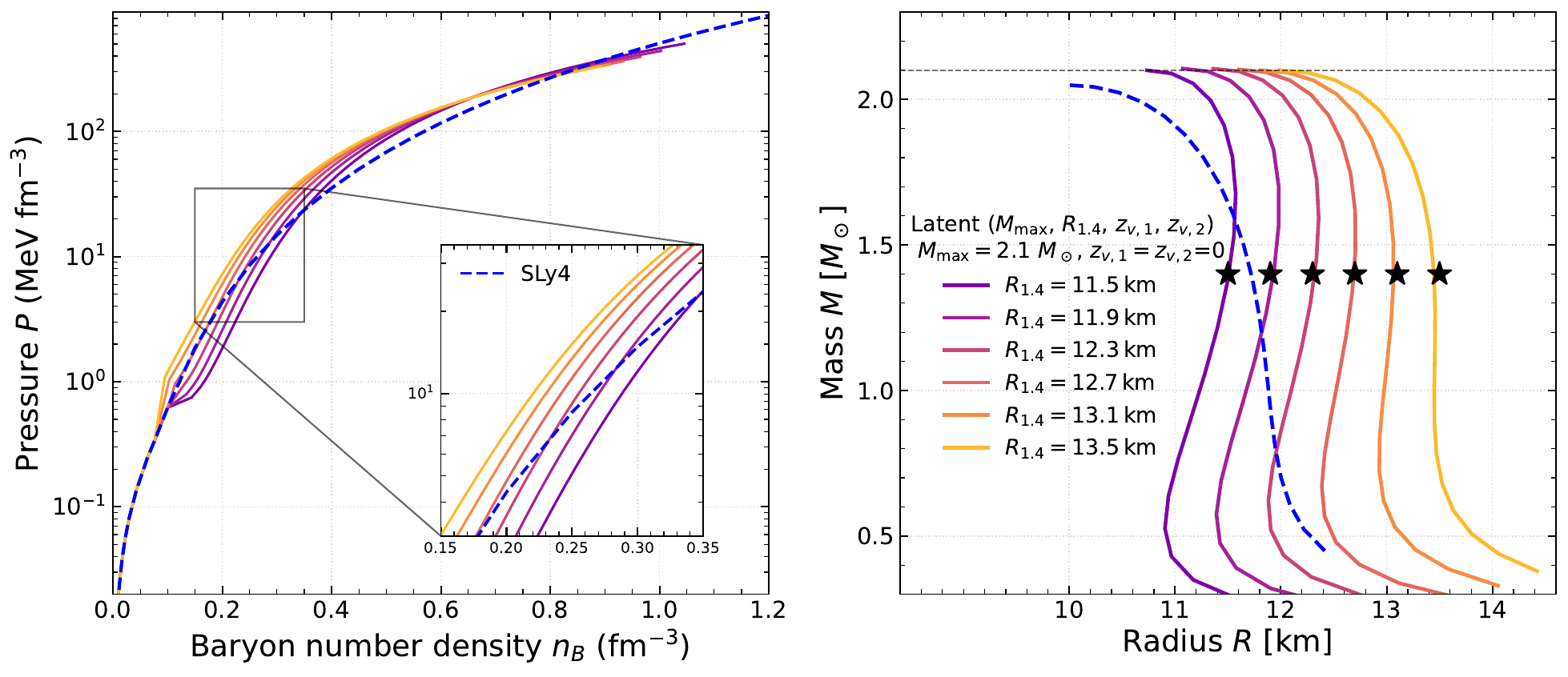}
    \includegraphics[width=0.8\columnwidth]{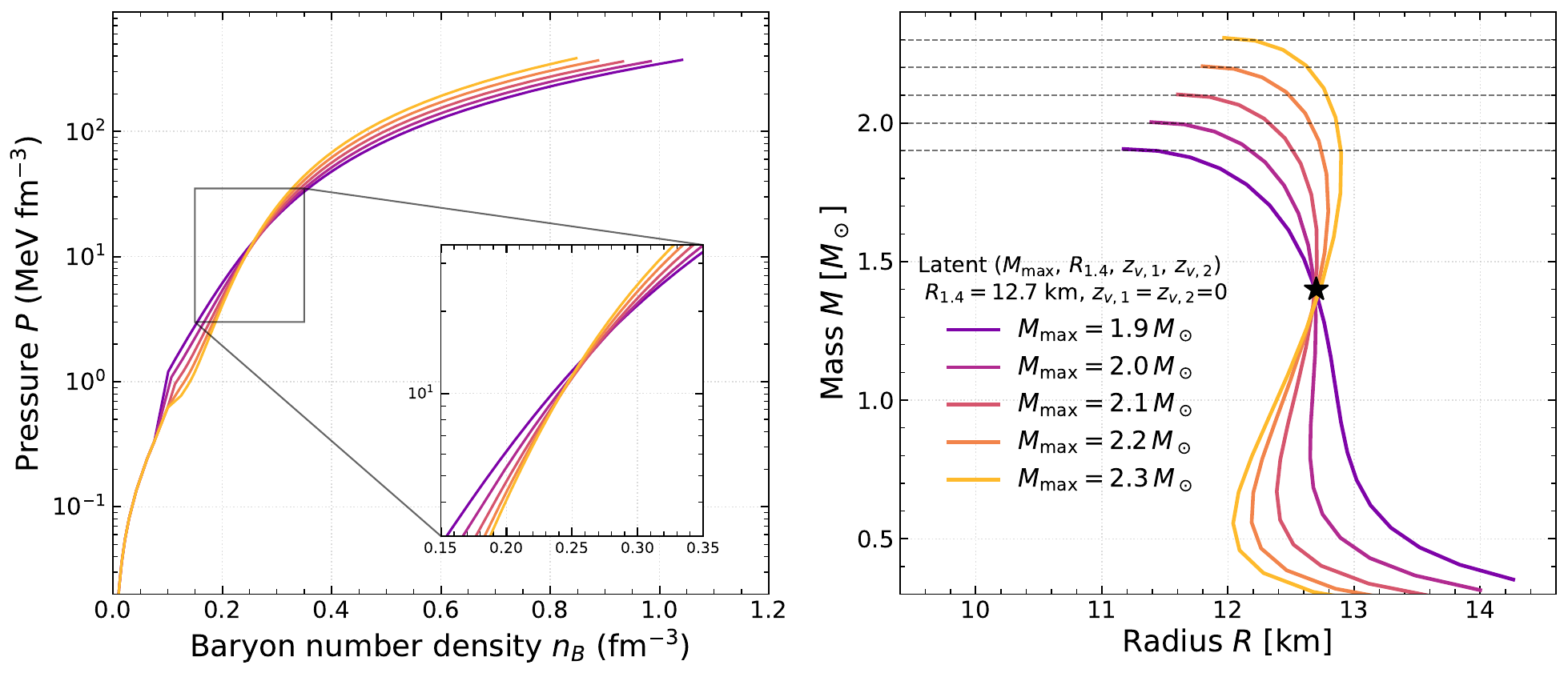}
    \caption{{Same as Fig. \ref{fig:EOS_sensitivity_Mmax_P_vs_nB} and Fig. \ref{fig:EOS_sensitivity_R14_P_vs_nB_zoom} but for RMF dataset.\label{fig:EOS_sensitivity_rmf}}}
\end{figure}
{Figure~\ref{fig:EOS_sensitivity_rmf} shows the sensitivity of the EOS and the corresponding MR relations to variations in $M_{\max}$ and $R_{1.4}$ while fixing the variational latent parameters at $z_{v,1}=z_{v,2}=0$. When $R_{1.4}$ is fixed, the PSL-VAE-generated EOSs show a similar crossing at intermediate baryon number densities around $n_B\approx0.25$ fm$^{-3}$ as in the Skyrme case. A major difference is that, for PSL-VAE-generated EOSs with the same $M_{\max}$ and $R_{1.4}$, the RMF-trained PSL-VAE tends to produce a larger $dR/dM$ around $M=1.4,M_\odot$ than the Skyrme-trained PSL-VAE. As a result, $R_{\max}$ is systematically larger for the RMF models than for the Skyrme models, suggesting that $R_{\max}$ could be an interesting additional supervised observable to consider in future applications.}

\begin{figure}[ht!]
    \centering
    \includegraphics[width=0.8\columnwidth]{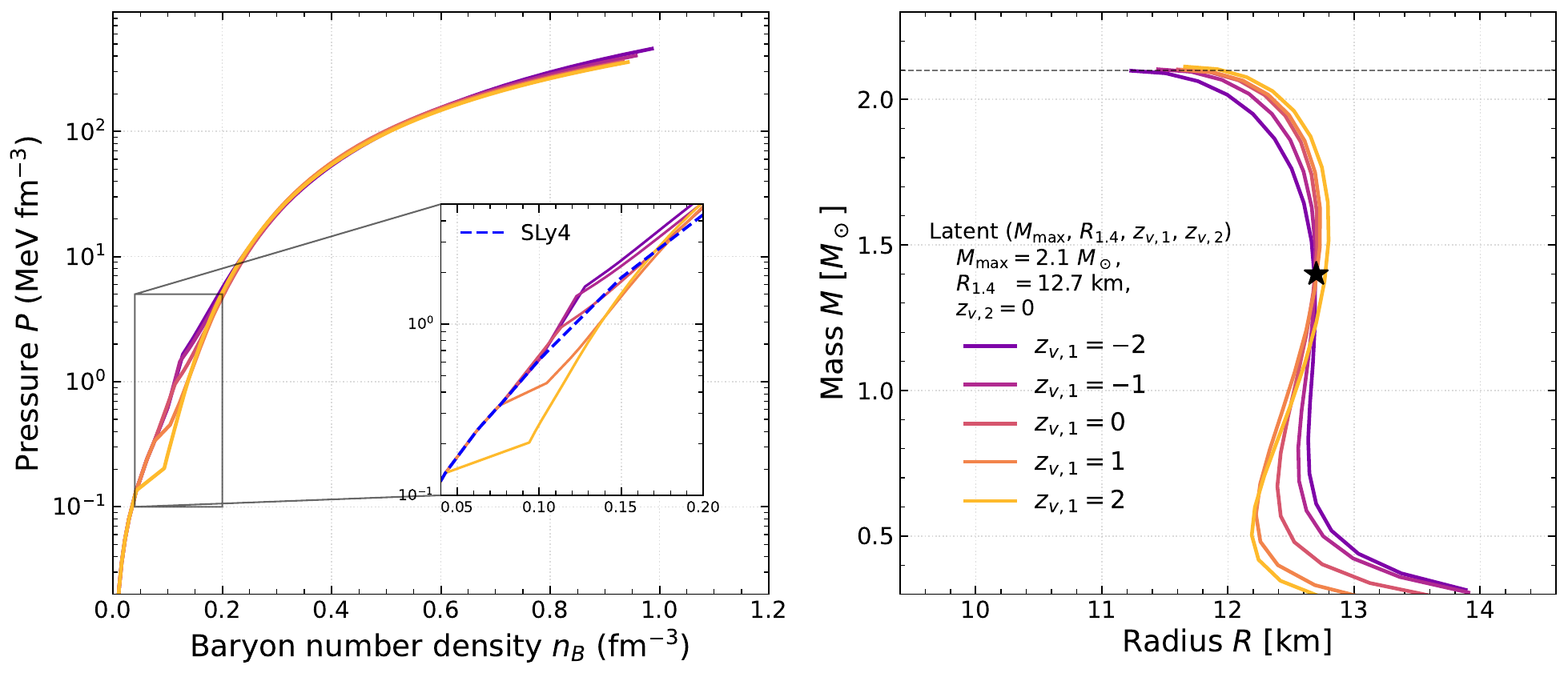}
    \includegraphics[width=0.8\columnwidth]{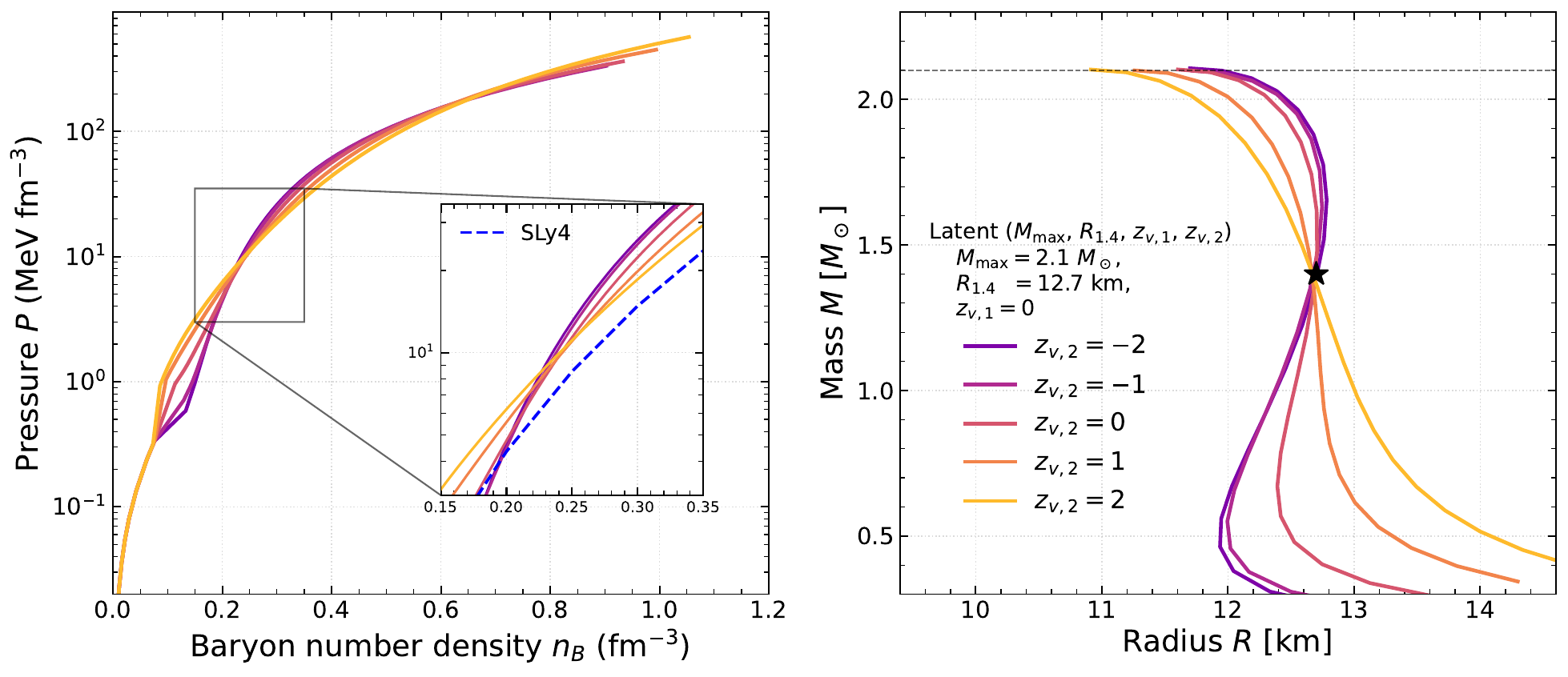}
    \caption{{Same as Fig.~\ref{fig:EOS_sensitivity_variational}, but for the RMF dataset. The selected PSL-VAE network for the RMF dataset has two variational latent parameters, $z_{v,0}$ and $z_{v,1}$, which are varied separately.}\label{fig:EOS_sensitivity_variational_rmf}} 
\end{figure}
{Figure~\ref{fig:EOS_sensitivity_variational_rmf} shows the sensitivity of the EOS and the corresponding MR relations to the variational latent parameters $z_{v,1}$ and $z_{v,2}$ while fixing $M_{\max}=2.1\, M_\odot$ and $R_{1.4}=12.7$ km. We find that $z_{v,1}$ and $z_{v,2}$ control different features of the EOS. In this particular trained network, $z_{v,1}$ primarily modifies the low-density EOS close to the crust-core transition, whereas $z_{v,2}$ primarily modifies the pressure around twice saturation density while approximately preserving the pressure at $n_B\approx0.25$ fm$^{-3}$ and $n_B\approx0.65$ fm$^{-3}$. As emphasized previously for the Skyrme model, the particular combinations represented by $z_{v,1}$ and $z_{v,2}$, as well as their effects on the EOS, are not unique and can vary among independently trained PSL-VAE networks. Therefore, the behavior shown in Fig.~\ref{fig:EOS_sensitivity_variational_rmf} is specific to the selected PSL-VAE network, whereas networks trained to comparable accuracy exhibit consistent dependence on the supervised observables shown in Fig.~\ref{fig:EOS_sensitivity_rmf}.}


\printbibliography

\end{document}